\documentclass[twocolumn, hyperref]{aa}  

\usepackage{multicol}
\usepackage{url}
\usepackage{longtable}
\usepackage{graphicx}
\usepackage{adjustbox}
\usepackage{hyperref}
\usepackage{xcolor}

\hypersetup{
 colorlinks = true,
  linkcolor = red,
  anchorcolor = blue,
  citecolor = blue,
  filecolor = blue,
  urlcolor = blue
}
\usepackage{txfonts}
\usepackage[normalem]{ulem}
\usepackage{amssymb}
\usepackage{esvect}
\usepackage{natbib,twoopt}
\usepackage{float}
\usepackage{color}

\usepackage{xcolor}
\begin{document}

   \title{The large-scale structure around the Fornax-Eridanus Complex}
\titlerunning{LSS around the Fornax-Eridanus Complex}

   \author{M. A. Raj
          \inst{1}, 
          P. Awad \inst{1,2},
          R. F. Peletier \inst{1}, 
          R. Smith \inst{3}, 
          U. Kuchner \inst{4}, 
          R. van de Weygaert \inst{1}, 
          N. I. Libeskind \inst{5},
          M. Canducci \inst{6}, 
          P. Ti{\v{n}}o \inst{6},
           \and 
       K. Bunte \inst{2}
        }

\institute{
Kapteyn Astronomical Institute, University of Groningen, PO Box 800, 9700 AV Groningen, The Netherlands
\label{kapteyn}
\newline
\email{m.a.raj@rug.nl}
\and
Bernoulli Institute for Mathematics, Computer Science and Artificial Intelligence, University of Groningen, 9700AK Groningen, The Netherlands
\label{bernoulli}
\and 
Departamento de F\'isica, Universidad Técnica Federico Santa María, Avenida Vicuña Mackenna 3939, San Joaquín, Santiago de Chile \label{inst3}
\and
School of Physics, Astronomy, University of Nottingham, Nottingham NG7 2RD, United Kingdom 
\label{inst4}
\and 
Leibniz-Institute für Astrophysik Potsdam (AIP), An der Sternwarte 16, D-14482 Potsdam, Germany
\label{inst5}
\and
School of Computer Science, University of Birmingham, Birmingham B15 1TT, United Kingdom
 \label{inst6}
   }
   \date{Received ... ; accepted ...}

   \date{Received xxx; accepted yyy}

 
  \abstract
   {}
   {Our objectives are to map the filamentary network around the Fornax-Eridanus Complex and probe the influence of the local environment on galaxy morphology.}
   {We employ the novel machine-learning tool, 1-DREAM (1-Dimensional, Recovery, Extraction, and Analysis of Manifolds) to detect and model filaments around the Fornax cluster. We then use the morphology-density relation of galaxies to examine the variation in the galaxies' morphology with respect to their distance from the central axis of the detected filaments.} 
   {We detect 27 filaments that vary in length and galaxy-number density around the Fornax-Eridanus Complex. We find that 81\% of galaxies in our catalogue belong to filaments and 19\% of galaxies are located outside filaments. The filaments around the Fornax-Eridanus Complex showcase a variety of environments; some filaments encompass groups/clusters, while others are only inhabited by galaxies in pristine filamentary environments. In this context, we reveal a well-known structure -- the \textit{Fornax Wall}, that passes through the Dorado group, Fornax cluster, and Eridanus supergroup. With regard to the morphology of galaxies, we find that early-type galaxies (ETGs) populate high-density filaments and high-density regions of the Fornax Wall. Furthermore, the fraction of the ETG-population decreases as the distance to the central axis of the filament increases. The fraction of late-type galaxies (LTGs; 8\%) is lower than that of ETGs (12\%) at 0.5~Mpc/$h$ from the filament spine. Of the total galaxy population in filaments around the Fornax-Eridanus Complex, $\sim$ 7\% are ETGs and $\sim$ 24\% are LTGs located in pristine environments of filaments, while $\sim$ 27\% are ETGs and $\sim$ 42\% are LTGs in groups/clusters within filaments. Among the galaxies in the filamentary network around the Fornax-Eridanus Complex, 44\% of them belong to the Fornax Wall.}
   {This study reveals the Cosmic Web around the Fornax Cluster, which exhibits a variety of filamentary environments. With this, our research asserts that filamentary environments are heterogeneous in nature. When investigating the role of the environment on galaxy morphology, it is essential to consider both, the local number-density and a galaxy's proximity to the filament spine (\textit{filament core}). Within this framework, we ascribe the observed morphological segregation in the Fornax Wall to pre-processing of galaxies within groups embedded in it.}


  \titlerunning{LSS around Fornax-Eridanus}
   \authorrunning{M. A. Raj et al. }
   \maketitle
%

\section{Introduction}
The large-scale distribution of matter and galaxies forms a web-like structure called the \textit{Cosmic Web} \citep{Bond1996}. Early attempts to map the Universe \citep{joeveer78, Geller89, Bond1996, vdW08, vDW09, Tempel14, Courtois13, Cautun14} revealed an intricate pattern containing structures extending from a few megaparsecs to several hundreds of megaparsecs. In a $\Lambda$CDM paradigm, these structures form hierarchically under the influence of gravity \citep{White78}, and the \textit{Cosmic Web} is therefore an accumulation of \textit{galaxies, gas}, and \textit{dark matter} in dense clusters, connected by filaments and walls, and separated by voids. Several galaxy redshift surveys  \citep[e.g.][]{Colless01, 2mass, tegmark04, huchra12, Desi} have confirmed the web-like network of galaxies. Likewise, N-Body simulations \citep[e.g.][]{ Vogel14, springel05, schaye15} of the Cosmic Web and cosmic density field reconstruction \citep[e.g.][]{Kitaura13} have also displayed the spatial pattern of galaxies as well as dark matter \citep[e.g.][]{Hes13, Sorce16, Kitaura23, Malavasi23}.\par 
The densest parts of the Cosmic Web are groups and clusters (sometimes called nodes) which may be part of superclusters \citep{deVauc58, einasto03, einasto11, liivam12} of galaxies, and these superclusters may be embedded in walls \citep{einasto10, Einasto17}. In this context, \citet{einasto24} showed that very rich clusters are located in superclusters and luminous groups and clusters are within or near filaments. Supercluster-Complexes are structures of several nearly connected, very rich superclusters, such as the Sloan Great Wall \citep{Gott05} and the BOSS Great Wall  \citep{lietzen16}. Additionally, superclusters can form vast planes spanning hundreds of megaparsecs, like the Local Supercluster Plane or the Dominant Supercluster Plane \citep{einasto1983, einasto97, peebles23}. Studies pertaining to superclusters \citep[e.g.][and references therein]{araya09, Chon14, tully14, kraan17, Haines18, Einasto21, Bag23, hatam23} showed that they are distinct in terms of their dynamical state, galaxy populations, and size. For example, \citet{Venturi22} detected minor mergers of galaxy-groups within the Shapely \citep{shapley} supercluster, the largest known gravitationally bound system. \par
Galaxies flow from low-density regions to high-density regions and therefore experience various environments throughout their lifetime. As such, these environments vary in density, dynamical state, and composition. The impact of the local density on the morphology of galaxies has been well-established by the morphology-density relation by \citet{dressler1980}, but understanding the additional effect of the topology or geometry of the physical environment of the Cosmic Web components is currently a topic of active research. Specifically, mechanisms acting in cluster and group environments have been shown to affect the star-formation of their member galaxies \citep[e.g.][]{gunn72, kauffmann04,Boselli06,bamford09, benedetta15, brown23}.  There is substantial observational evidence of the morphology-density relation in clusters and groups  \citep[][and references therein]{simona23, marasco23}, wherein early-type galaxies (ETGs) are commonly found in high-density regions and late-type galaxies (LTGs) are prevalent in low-density regions. \citet{einasto87} were the first to show that the morphology-density relation extends from rich clusters to poor groups as well as isolated galaxies. \citet{einasto92} also concluded that ETGs either belong to clusters or that they reside in small groups within filaments. However, the consequence of these occurrences is still a debate. In particular, it is still unclear whether density is a predominant cause for the formation of ETGs or that these galaxies formed at an earlier epoch. With regard to the low-stellar mass regime ($M_* = 10 ^7-10^9 M_{\odot}$), research \citep[e.g.][to name a few]{Roman2017b, Benavides23, For23} in the past decade has revealed evidence of the role of the environment on dwarf galaxies, with particular focus on their formation channels. The question persists as to the extent of environmental influence on the formation and evolution of dwarf galaxies. In fact, we are yet to investigate them in cosmic environments. However, the role of cosmic environments on the evolution of galaxies ($M_* \geq 10^9 M_{\odot}$) has been corroborated extensively, with some indication of overlapping densities in filaments, walls, and clusters  \citep[see][]{Cautun14, Libeskind2018}. In this context, the extent of this relation to large-scale structures \citep[e.g.][]{kraljic18, einasto22, espinosa23} is also an ongoing research topic. As such, there is evidence of a strong correlation of galaxies' properties such as colour, stellar mass, and star-formation, with regard to the distance to their filament spine \citep[e.g.][]{chen17, kraljic18, casti22, bulichi23}. Several theoretical explanations exist to explain this observed trend. For example, \citet{calvo19} proposed that star-formation of galaxies is quenched when they are detached from their primordial filaments.\par
Dark matter and baryonic matter are fed into clusters and groups through cosmic filaments \citep[e.g.][]{calvo07, Cautun12, espinosa23, Cornwell2024}, and within these filaments, galaxies start to experience ``pre-processing'' \citep{zab98, Balogh2000, fujita04}. This term has a broad range of definitions in the literature, but overall, it includes all processes affecting the properties of galaxies before cluster-infall. There have been extraordinary evidences of pre-processing of galaxies, not just in groups accreting onto clusters \citep[e.g.][]{kleiner21, Bidaran22, Lopes2024} but also in filaments \citep[e.g][]{Cybul14, sarron19, Chung21}. \citet{Kuchner2022} also demonstrated the heterogeneity of filament environments and the various pathways galaxies take before cluster-infall. \par
To investigate mass-accretion in the Cosmic Web and the impact of the large-scale environment on galaxy evolution, we require robust tools to classify large-scale structures. \citet{Libeskind2018} discussed the functionality of several such techniques. Some prominent examples are DisPerSE \citep{sousbie11}, Bisous model \citep{Tempel14}, Multi-scale morphology filter/NEXUS \citep{calvo07, Cautun13}, V-web: velocity shear tensor \citep{libeskind14}, or T-REx (Tree-based Ridge Extractor; \citealt{Bonnaire20, aghanim24}). Based on the objectives for classification and the methodologies in comparing the extracted components of the Cosmic Web, \citet{Libeskind2018} found that these different techniques yielded diverse results on the properties of the Cosmic Web despite several similarities. Another caveat is that the filamentary structures defined theoretically and observationally are those traced by massive galaxies, and this is due to the detection limit of current galaxy surveys \citep[e.g.][]{Cautun14}. Hence, Cosmic Web classification methods should be reliable in detecting tenuous structures and optimised to include dwarf galaxies.\par
Our analysis is based on the novel toolbox for analysing the Cosmic Web, called 1-DREAM\footnote{1-DREAM is publicly available at \url{https://git.lwp.rug.nl/cs.projects/1DREAM}}, an acronym for \textit{1-Dimensional Recovery, Extraction, and Analysis of Manifolds.}  1-DREAM was first introduced by \citet{Canducci2022} and followed by this, \citet{Awad2023}  demonstrated its application and reliability on N-body simulations.  In brief, 1-DREAM comprises five different Machine Learning (ML) algorithms  that serve the extraction and analysis of multi-dimensional structures within astronomical datasets. The functions performed by the algorithms range from detecting cosmic structures across varying densities  and distinguishing the different environments of the Cosmic Web, to finding the central spines of the detected filaments and probabilistic modelling of the distribution of matter surrounding them. In addition to giving a detailed description of the functionality of 1-DREAM, \citet{Awad2023} applied it on the same cosmological N-body simulation used by \citet{Libeskind2018} as a common data set to compare it to many Cosmic Web-tracing methodologies. The toolbox produced comparable results to the state-of-the-art methods when separating the simulated particles between each Cosmic Web environment. Notably, \citet{Awad2023} showed that reducing the number of particles comprising filaments in cosmological simulations produced minimal change in the central axes retrieved for the same filaments. Such stability is not found in widely used Cosmic Web-tracing algorithms such as DisPerSE, where the down-sampling of data tends to cause visible changes in the retrieved axes. This advantage is crucial, considering the sparsity of galaxies per unit volume within observations in comparison to cosmological simulations of the Cosmic Web. \citet{Awad2023} also confirmed that the filament spines found by 1-DREAM tend to trace accurate paths that become highly stable references which in turn allows accurate measurements of the properties of galaxies as a function of their distances to the axes. Finally, \citet{Awad2023} showed that 1-DREAM produces less false positive or negative detection of filaments when compared to DisPerSE. These advantages have motivated the use of 1-DREAM in our investigation of the large-scale structure of the Fornax-Eridanus Supercluster. The research presented here is both a first used-case for the 1-DREAM toolbox on observational data \textit{and} the largest systematic analysis of galaxy pre-processing in the extended region of the Fornax-Eridanus Complex.\par
The Fornax-Eridanus Complex is at distance of 20~Mpc\footnote{The Fornax cluster (GLON = $236.7163^\circ$; GLAT $= -53.6356^{\circ}$) is located in the southern Galactic hemisphere (for reference, our Milky Way is at GLON $= 359.9443^{\circ}$ and GLAT$= -0.0461^{\circ}$) and the Virgo cluster (GLON $= 283.7775^\circ$; GLAT $= 74.4912^\circ$) is in the north Galactic hemisphere (see Fig.~12 and Sect.~10 of \citealt{Kourkchi2017}).}. We aim to probe the Cosmic Web around this region and examine the role of the filamentary environment on the morphology of galaxies. The research presented in this article is also motivated by the wealth of imaging data currently available (and with upcoming data from the Euclid Survey; \citealt{euclid2022}) for the region around the Fornax cluster. This will allow us to investigate the influence of the large-scale environment on several observational properties of galaxies. Below, we give a literature review of the Fornax-Eridanus Complex. \par
 The Southern Supercluster is a long strand of galaxies from Cetus to the Dorado group, passing through Fornax, Eridanus, and Horologium \citep{devauc53, deVauc56}. \citet{Fairall94} described this supercluster as a ``pancake structure'' (flattened) that resembled a \textit{Great Wall} and called it the Fornax Wall. Within this ensemble of galaxies, \citet{nasonova2011} and \citet{Makarov2011} found that several groups were merging in the Fornax and Eridanus regions \citep[see also][]{Brough06} albeit a majority of them are yet to reach dynamical equilibrium. The major systems forming the Fornax-Eridanus Complex are the Fornax cluster (NGC~1399) and its infalling subgroup (NGC~1316), the Eridanus supergroup (NGC~1302, NGC~1332, NGC~1395, NGC~1398, NGC~1407; see \citealt{Brough06}), and the Dorado group (NGC~1553, NGC~1672). \citet{nasonova2011} found that smaller systems such as galaxy pairs and triplets were merging with the Fornax cluster. Similarly, the groups within the Eridanus cloud will eventually merge to form a cluster with NGC~1407 as the Brightest Cluster Galaxy. The Cetus-Aries cloud \citep[see][]{devauc75, Tully1987a} comprises several groups, some of which belong to the Cetus I cloud (NGC~1052, NGC~1068) and the Cetus II cloud (NGC~0584, NGC~0681). On larger scales, the Southern supercluster is part of the Southern Supercluster Strand, which is a filament connecting to the Centaurus cluster \citep{Courtois13, tully14}. The other branch of this Strand is the Telescopium-Grus Cloud, a low-density filament, which connects to the Southern Supercluster through the Cetus-Aries Cloud \citep{tully87}. Collectively, the two branches of the Southern Supercluster Strand are called SSCa (Fornax-Eridanus-Dorado) and SSCb \citep[see][]{Courtois13}; the Southern Supercluster Strand is in turn part of the Laniakea Supercluster \citep{tully14}.\par
The layout of this paper is as follows: We explain the dataset and method we employ in Sects.~\ref{sect:data} and \ref{methods} respectively.  In Sect.~\ref{Filaments FS},  we present our results of the detected filaments, followed by their analysis in Sect.~\ref{sec:Morphology}. We then discuss the role of the large-scale environment on galaxy morphology in Sect.~\ref{sect:disc}. Finally, we summarise our results, draw conclusions of the same, and portray our future perspectives of this study in Sect.~\ref{sect: summary}. We provide supplementary information in Appendix~\ref{appendix}. \par
\begin{figure*}[h]
   \centering
   \includegraphics[width=\textwidth]{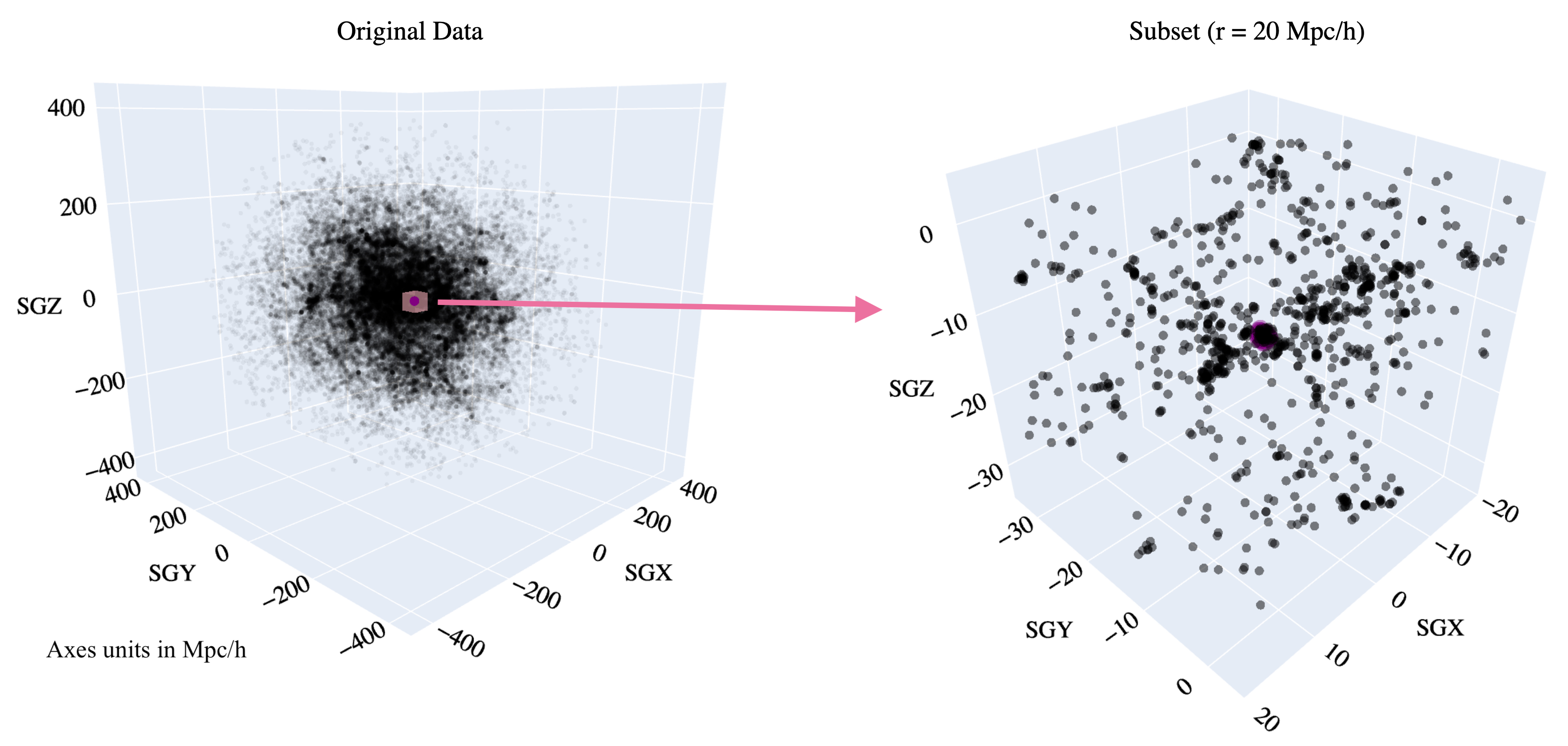}
      \caption{Three-dimensional distribution of galaxies from the group catalogue by \citet{Tempel2016a} represented in supergalactic coordinates (SGX, SGY, SGZ). The selected data with $r=20$ Mpc/$h$ centred on the Fornax cluster is highlighted in a pink mesh on the \textit{left} panel and the zoomed-in cube is displayed on the \textit{right} panel.}
           
         \label{Data_sub}
   \end{figure*}  
\section{Data} \label{sect:data}
Our analysis is based on the galaxy-group catalogue by \citet{Tempel2016a} (hereafter T16a) which was constructed using data from the Extragalactic Distance Database \citep[EDD,][]{Tully09}. The dataset used by T16a is primarily based on the 2MASS Redshift Survey data \citep{huchra12}. To complement this, T16a also included redshift sources from the 2M++\footnote{The 2M++ survey is based on data from the Two-Micron-All-Sky \citep{2mass} Source Catalogue and the 6dF Galaxy Survey \citep{Jones09}.} survey \citep{Lavaux11}, as well as CosmicFlows-2 \citep{Tully13}.  In brief, T16a employed the friends-of-friends algorithm in cosmology (see \citealt{Tempel14} for procedure), which was introduced by \citet{zeldovich82} and \citet{huchra82}. Then, they used a Multivariate Gaussian Mixture Modelling approach to identify substructures within galaxy-groups. Finally, T16a removed galaxies which were not bound to groups based on a group's virial radius and escape velocity. Additionally, the redshift space distortions for groups were suppressed as part of their method (see also \citealt{Tempel14} for a detailed description about the Finger of God correction)\footnote{This correction is an essential step to perform, otherwise the Fingers of God will appear as over-densities in the data, and so will be mistaken by our methodologies as real filaments.}. The catalogue by T16a comprises 78378 galaxies of which, 43480 are brighter than $K_S < 11.75 $ mag (from the 2MASS Redshift Survey), 3627 do not have measured $K_s$ magnitudes (CosmicFlows-2), and the remaining from 2M++ go down to $K_s < 12.5 $ mag. This catalogue is complete for galaxies with $K_s < 11.5$ mag, which corresponds to log$_{10} (M_{*}/M_{\odot}) = 8.40$ at a distance of 10 Mpc \citep{dalya18, Ducoin20}. Therefore, this catalogue includes the brightest dwarf galaxies as well as more massive galaxies. Comparing to the Local Group, this would include galaxies such as the Large Magellanic Cloud, M33, M31, and our Milky Way. As the research presented in this article focuses on the region surrounding the Fornax-Eridanus supercluster, we extracted a subset of $r = 20 $~Mpc$/h$, which we show in Fig.~\ref{Data_sub}; the total number of galaxies within this subset is 933. Hereafter, we refer to this subset unless the original dataset is mentioned explicitly. 
\section{Method} \label{methods}
We employ 1-DREAM to extract filamentary structures around the Fornax-Eridanus Complex. In this section, we summarise 1-DREAM, its usage, and the procedure we adopted to extract filaments from real data. 
\subsection {1-DREAM}
Here, we list the five algorithms within the toolbox, the acronyms used to refer to each of them, and their corresponding functions (for a detailed mathematical description, see \citealt{Canducci2022} and \citealt{Awad2023}):
\begin{enumerate}
    \item Locally Aligned Ant Technique (LAAT): It initiates a random walk through the data and distributes an artificially defined quantity termed the “pheromone” in reliance on the concept of Ant Colony dynamics. The random walk is guided through the stipulation that local jumps between the points are more likely to occur if these points are aligned along a low-dimensional spatial structure. Running LAAT allows the pheromone to be concentrated in regions of high density and regions aligned with locally detected manifolds or structures. The pheromone quantity therefore acts as a positive feedback mechanism which enhances the contrast between low-density and high-density regions, and can be used as a threshold parameter to separate regions of varying density \citep{Taghribi2023}.   
    \item Evolutionary Manifold Aligned Aware Agents (EM3A): This collapses the positions of data points belonging to a structure onto that structure's central axis. The resulting distribution of points traces the spines of the structures detected in the previous step  \citep{Mohammadi2020_EM3A}. We refer to this spine as the filament core.    
    \item Dimensionality Index (DimIndex): It distinguishes the underlying dimensionality of the local structures that the particles/points belong to, which in turn allows the separation of 1-dimensional, 2-dimensional, and 3-dimensional (3D) structures {\citep{Canducci_2022_Prob}}.   
    \item Multi-manifold Crawling (MMCrawling): It partitions the data into a set of filaments represented by their central axis and the group of particles surrounding them. Through this step, the spines (\textit{filament core}) are also represented by graphs, that is, a series of nodes connected together by edges \citep{Canducci_2022_Prob}.    
    \item Stream Generative Topographic Mapping (SGTM): This models the points belonging to each structure using a constrained Gaussian Mixture Model aligned along that structure. The purpose of this probabilistic modelling is to relax the idea that a structure begins and ends at a definite position. The model replaces this concept with a likelihood to belong to a structure instead \citep{Canducci_2022_Prob}.   
\end{enumerate}
\begin{figure*}[h]
   \centering
   \includegraphics[width=\textwidth]{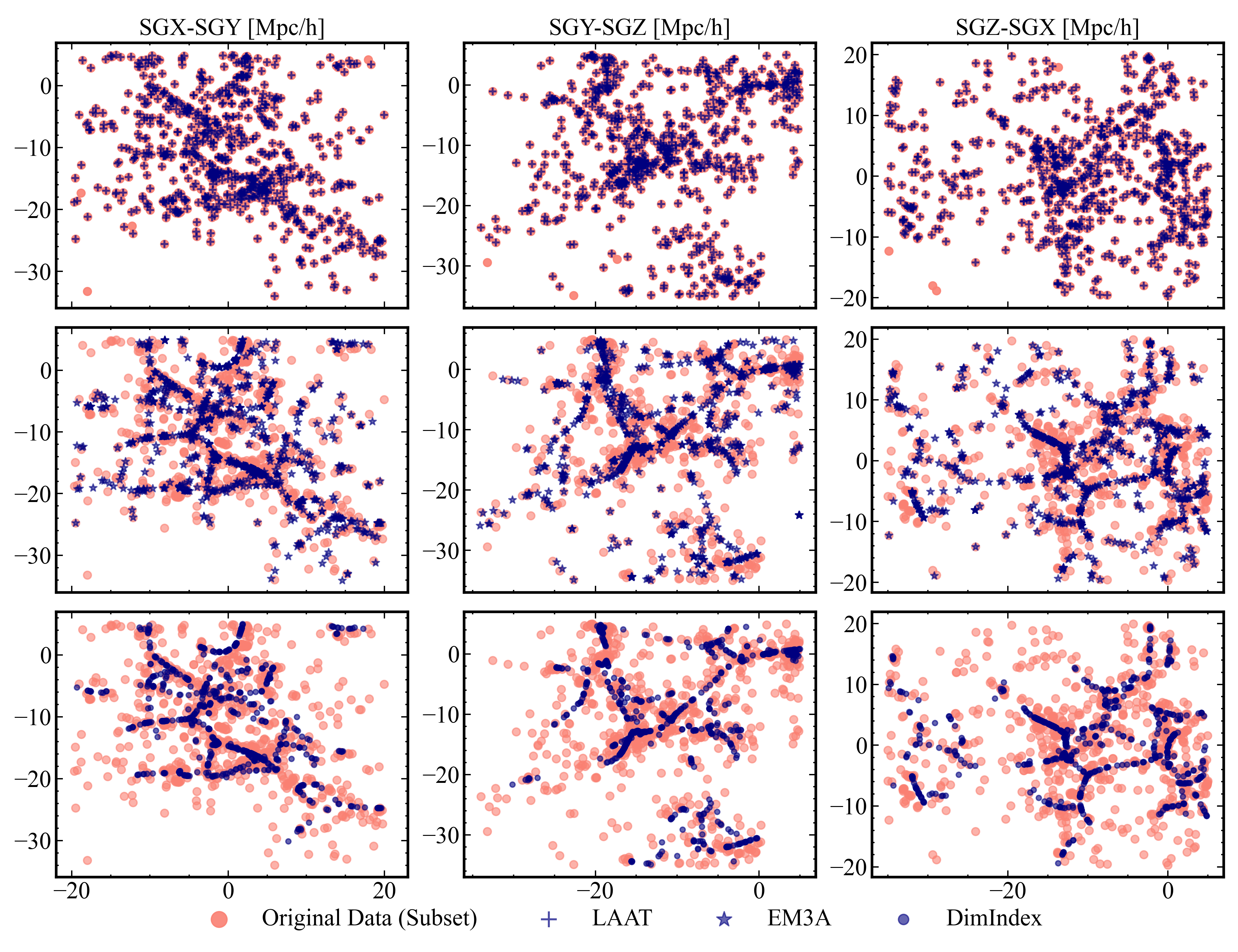}
      \caption{Results of LAAT, EM3A, and DimIndex algorithms applied to our data. The first row depicts the structures highlighted by LAAT compared to the original distribution of galaxies. The second row shows the axes enhanced by EM3A. The third row shows the remaining 1-dimensional structures found by DimIndex. The results of each algorithm are drawn in blue and compared to the original distribution of data drawn in orange. The columns from left to right represent projections along the $x-y$, $y-z$, and $z-x$ planes.}
           
         \label{Data_MBMS}
   \end{figure*}
\par In this particular application, the data points on which the algorithms run  are the individual galaxies in our sample. With regard to our current data selection, these different algorithms can be used consecutively in the following way: LAAT is used to highlight galaxies that belong to local structures present in the subset and allows their extraction from the data. EM3A can then be used to find the central axis running through the detected structures. DimIndex then separates the galaxies between those belonging to large clusters, walls, filaments, and voids. In this way, we isolate all detected filaments within the subset. MMCrawling is then used to create a catalogue of all central axes of the filaments as well as the galaxies surrounding them, up to a given distance. In the creation of this catalogue, MMCrawling represents the axes as graph objects as explained above. The subsequent step would be to use SGTM to regularise the position of the embedded graph nodes and model the distribution of galaxies. Due to the probabilistic nature of SGTM, a densely-sampled manifold is essential to train and obtain Maximum Likelihood Estimates of its parameters. However, the sparsity of the subset at hand is not adequate to sample the filaments. Thus, in order to avoid over-fitting the model parameters of the available data, we did not use SGTM as part of the analysis in this research. When working with simulations that tend to be more populated, we recommend the usage of SGTM as a final step of applying 1-DREAM.

\subsection{1-DREAM usage and optimised-Crawling on real data}
All algorithms within 1-DREAM rely on the idea of local neighbourhoods and localised Principal Component Analysis (PCA). In this context, local neighbourhoods are defined as spherical regions of a given radius $r$ centred on each galaxy. Through PCA, we infer the local eigen-directions in which the galaxies are distributed within their neighbourhood (defined by the aforementioned radius $r$). Within the toolbox, performing PCA in local neighbourhoods is the central step towards the intended function of each of its algorithms, and the size of the neighbourhoods plays a role in steering the PCA estimates. Adopting a very small radius acts as a zoomed-in perspective on the structures and might not include enough galaxies to infer correct information about their morphology. It would also lead to filtering out more than necessary galaxies as they get wrongly labelled as noise and not part of local structures. Similarly, choosing a neighbourhood radius that is larger than necessary leads to the inclusion of galaxies from multiple structures within the same neighbourhood and as a consequence, will bias the eigen-direction estimation and the conclusions regarding the morphology of the structures.\par
\begin{figure*}
   \centering
   \includegraphics[width=\textwidth]{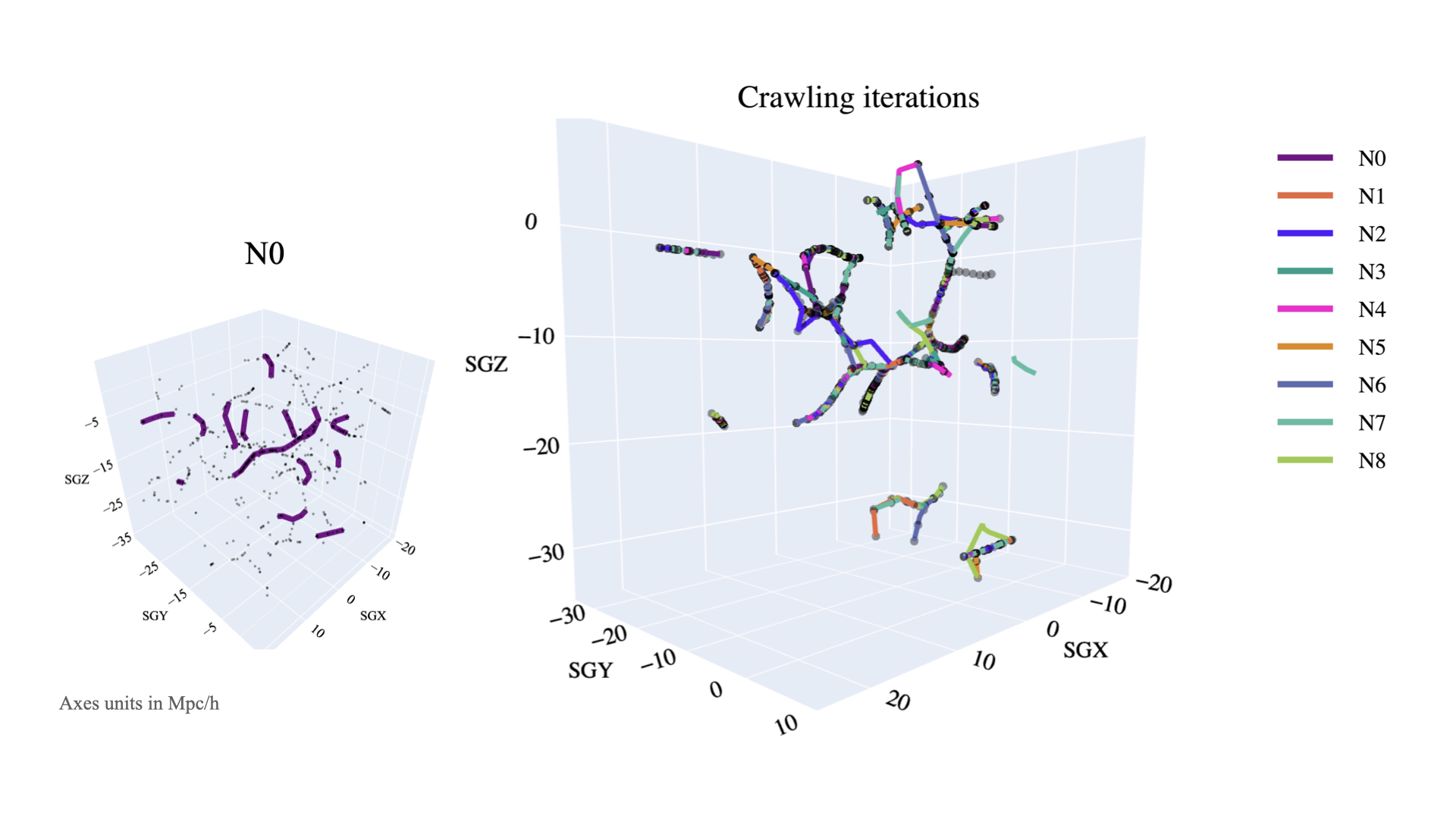}
      \caption{Results from MMCrawling on the filtered data. The distribution of galaxies in supergalactic coordinates (SGX, SGY, SGZ) is shown in black and the first iteration (N0) of MMCrawling is highlighted in purple on the \textit{left} panel. Results from all nine iterations from MMCrawling are represented in various colours on the \textit{right} panel. To visualise each iteration separately, see Appendix~\ref{Crawls_ite}. }
           
         \label{Crawls_all}
   \end{figure*}
The neighbourhood radius parameter is normally chosen to be a small multiple of the expected structure-thickness in the data. Given that filament and wall diameters range from 1--7~Mpc/$h$ \citep{Cautun14}, a suitable neighbourhood radius should lie within this range. Additionally, given the sparsity of the data, we expect to use a larger radius than what would be used for simulations to capture as much directional information as possible. Through experimentation, we found that a neighbourhood radius of 3~Mpc/$h$ produces the highest density contrast between the local structures and under-dense regions. It is also the least possible radius to use without visible structures getting eliminated by LAAT; this condition is crucial for recovering hidden structures within the region of interest. Therefore, we  consistently adopted this radius throughout the application of 1-DREAM to our data. In order to preserve the structures which extend beyond our region of interest, we ran LAAT and EM3A on the entire (original) T16a dataset and from there, extracted the same subset, a data cube of side length of 40~Mpc/$h$, centred on the Fornax cluster. The particles which survived the threshold of LAAT are shown in the first row of Fig.~\ref{Data_MBMS}. With EM3A, we recovered the central axes of the detected structures using the same value of the neighbourhood radius (3~Mpc/$h$). The resulting central axes are shown in the second row of Fig.~\ref{Data_MBMS}.  As our analysis focuses on cosmic filaments, we used DimIndex to extract structures that have a dimensionality of 1, that is, 1-dimensional structures. We depict the results of these three algorithms along projections on different planes in Fig.~\ref{Data_MBMS}, with each row highlighting LAAT, EM3A, and DimIndex. We also provide a 3D visualisation of these results in Appendix~\ref{LAAT_MBMS}.\par
We then modelled the axes of the recovered structures using MMCrawling. This algorithm operates by randomly choosing a seed point that lies on the recovered central axes from the previous steps of 1-DREAM.  It then crawls on both sides of the 1-dimensional manifold guided by the local PCA until it reaches the end of the filaments (refer to \citealt{Canducci2022} for the definition used for determining the end of a filament). \par 
The graph representing the first randomly chosen filament is then saved and the procedure is initiated again on the remaining number of galaxies, each time recovering an embedded graph object modelling the central axis of a given filament. The graphs as well as the galaxies that lie within neighbourhoods of size $r$ centred on the nodes are catalogued. At the end of a complete MMCrawling run, we obtained a catalogue of graphs that model the central axis of the filaments within the data and for each graph, a set of galaxies that lie within a distance $r$ from the corresponding filament core.
\begin{figure*}[!htbp]
   \centering
   \includegraphics[scale=0.41]{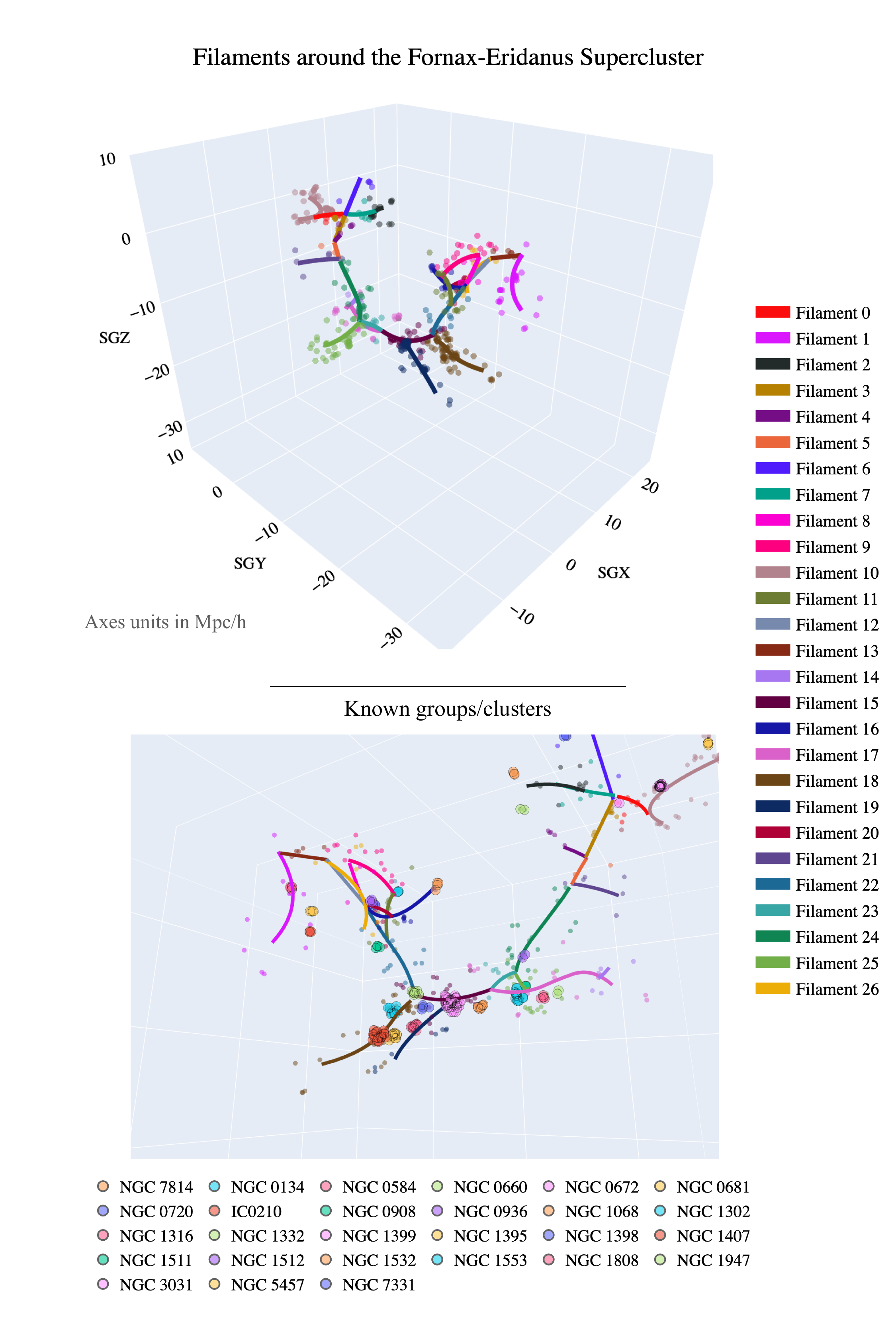}
      \caption{Filaments around the Fornax-Eridanus Complex. The distribution of galaxies belonging to 27 filament cores, displayed in supergalactic coordinates (SGX, SGY, SGZ), are represented in different colours on the \textit{top} and \textit{lower} panels. Each filament core is highlighted in the same colour as its member galaxies. Known groups/clusters from the group catalogues by \citet{Makarov2011} and \citet{Kourkchi2017}, which are part of the 27 filaments are also indicated in the \textit{lower} panel.}   
        \label{Filaments_fig}
\end{figure*}

\begin{table*}[h!]
\caption{Filaments detected around the Fornax-Eridanus Complex}              
 \label{tab:my_table1}    
\centering                                      
\begin{tabular}{lccc}          
\hline\hline                        
Filament name & Length of filament & $N_{gal}$  & Known groups/ clusters \\
  & Mpc/$h$ &  &  \\    
  (1)&(2)&(3)&(4)\\
 \hline\hline
   \textit{Filament 0} & 5.79 & 9 & NGC~0672\\
\textit{Filament 1} & 4.73 & 24 & IC~0210, NGC~0584, NGC~0681, \\
\textit{Filament 2} &4.27 & 16 &NGC~0660, NGC~7814\\
\textit{Filament 3} &7.3 & 9 & -\\
\textit{Filament 4} &1.74 &  5 & - \\
\textit{Filament 5} &1.59 & 2 & -\\
\textit{Filament 6} & 3.63 & 5 & NGC~7331\\
\textit{Filament 7} &4.69 & 5 & -\\
\textit{Filament 8} &1.5 & 6 & NGC~0720\\
\textit{Filament 9} & 12.41&  14  & -\\
\textit{Filament 10} &10.48 & 63 & NGC~3031, NGC~5457\\
\textit{Filament 11} & 7.8 & 20 &  NGC~0134, NGC~0908 \\
\textit{Filament 12} & 6.94&  1& - \\
\textit{Filament 13} &4.08&  5& - \\
\textit{Filament 14} &4.04&  5 &-\\
\textit{Filament 15} &8.81&  43&   NGC~1398, NGC~1399, NGC~1532\\
\textit{Filament 16}  &6.08 & 12 & NGC~1068\\
\textit{Filament 17}  &9.91&  20 & NGC~1808\\
\textit{Filament 18}  &7.95 & 76& NGC~1302, NGC~1332, NGC~1395, NGC~1407\\
\textit{Filament 19}  &5.95 & 50 & NGC~1316, NGC~1399\\
\textit{Filament 20} & 3.52 & 7& NGC~0936\\
\textit{Filament 21} &6.9 & 6& -\\
\textit{Filament 22}  &6.61 & 10 & - \\
\textit{Filament 23}  &4.18  &6& -\\
\textit{Filament 24}  &6.45 & 20 & NGC~1512\\
\textit{Filament 25}  &10.35 & 41& NGC~1511, NGC~1553, NGC~1947\\
\textit{Filament 26} & 4.05&  6& -\\

\hline
\end{tabular}
\tablefoot{Column 1 -- Filament identification number; Column  2 -- Length of filament in Mpc/$h$ ; Column 3 --  Number of galaxies within $r=3$ Mpc/$h$ of the filament spine; Column 4 -- Known groups/clusters from the group catalogues by \citet{Makarov2011} and \citet{Kourkchi2017}.} 
 \end{table*}
 
In terms of parameters, we adopted the same neighbourhood size of radius  $r = 3$~Mpc/$h$ and a distance between projected graph nodes of $\beta \times r = 1.5$~Mpc/$h$ where $\beta=0.5$. We chose the latter parameter to create the most reliable representations of the curvature of the filaments. Using a large $\beta$ could lead to large jumps while crawling along the central axes to create the graphs and therefore, will not follow the length of the filaments smoothly, especially in regions where curvature is present. After applying MMCrawling once, we obtained a set of graphs which represent the modelled filaments shown in the left panel of Fig.~\ref{Crawls_all}. \citet{Awad2023} discussed the effect of the stochastic nature of this algorithm on its results and showed that differences occurring as a result of varying the random starting point of crawling is minimal but can become more likely when applied to filaments of low density. Therefore, given the sparsity of our observational data set, we expect to find differences in the modelled central axes when running MMCrawling several times (each time varying the randomly chosen starting position). To overcome this limitation and to ensure that the filaments we extracted were real and reliable, we ran Crawling numerous times until the graphs of the filaments we obtained overlapped with each other. Regions of overlapping graphs therefore indicate an agreement between the different runs and thus, point at reliable structures within the data. The resulting graphs from nine iterations are shown in Fig.~\ref{Crawls_all} (right panel). We provide the separate results of each iteration in Appendix~\ref{Crawls_ite}).

We matched the coordinates of the nodes common between the graphs from all iterations. The segments connecting these nodes are further interpolated using cubic splines and then up-sampled to facilitate subsequent measurements such as the distances between the galaxies and their corresponding filament spine. The parameters mentioned in this sections are the only modified part from the suggested default-settings of all algorithms; these can be found in Table~\ref{tab:my_label}. As we focus only on the filaments connected to the Fornax-Eridanus Complex, the filaments we present in this article pertain only to this region. Within this region, we detected 27 filaments, which we list in Table~\ref{tab:my_table1}.

After up-sampling the modelled central axis for each retrieved filament, we repeated the search for galaxies near the identified cenral axes of filaments. For this, we used the {\ttfamily search\_around\_3D} function from Python's {\ttfamily astropy} module to find the nearest galaxies within a radius of 3~Mpc/$h$ of each filament core. With this function, we also obtained the 3D Cartesian distance of each galaxy to its filament core.  

\begin{figure*}[!htbp]
   \includegraphics[width=\textwidth]{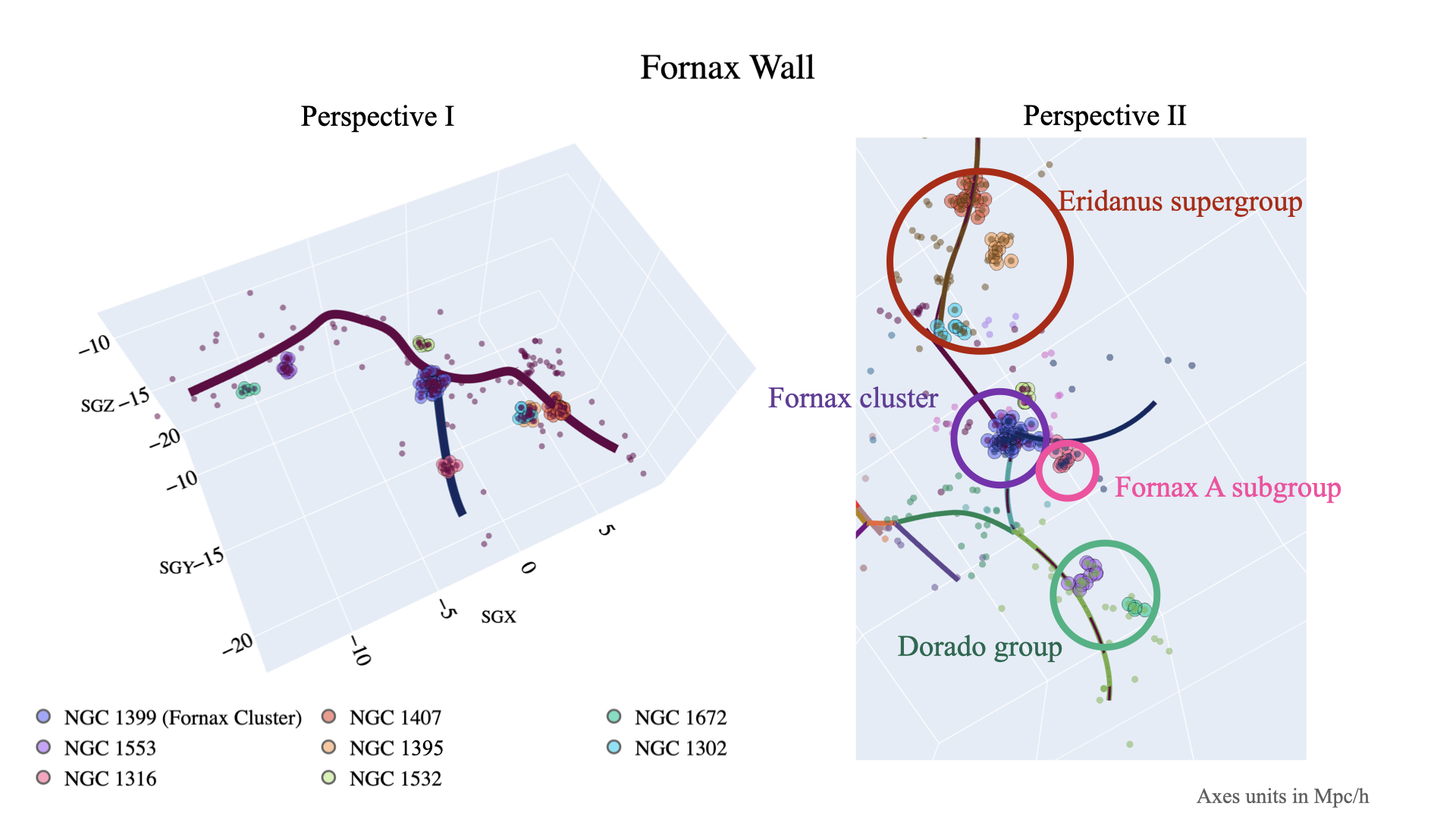}
      \caption{Central axis of the Fornax Wall and its constituent galaxies. Two perspectives of the Fornax Wall are shown in each panel and all known groups/clusters are indicated in different colours. Perspective I (\textit{left}) showcases the distribution of galaxies along the Fornax Wall as well as Filament~19, in supergalactic coordinates (SGX, SGY, SGZ), while Perspective II (\textit{right}) is a snapshot of the same along the $x-y$ planes. The prominent structures -- Fornax cluster, Fornax~A subgroup, Eridanus supergroup, and the Dorado group are also indicated in Perspective II  (\textit{right}). }         
         \label{FS}
   \end{figure*}

\section{Filaments around the Fornax-Eridanus Complex} \label{Filaments FS}
In this section, we characterise the Fornax-Eridanus Complex by connecting our findings from 1-DREAM with known groups/clusters from the literature. We indicate the central axis of all 27 filaments and their member galaxies within our data cube in Fig.~\ref{Filaments_fig} (top panel). The number of galaxies within a neighbourhood radius of 3~Mpc/$h$ of these filament cores is 486. Of the extracted subset at hand, 177 galaxies are located outside filaments (beyond 3~Mpc/$h$ of the filament spines). 

The filaments are connected to each other and form a coherent structure. This filamentary network is disconnected to filaments at the edge of the investigated volume. A possible explanation for this is the presence of fewer galaxies between filaments towards the edge and the filaments connected to the Fornax-Eridanus Complex. The remaining 270 galaxies from the extracted subset belong to these filaments that are not connected to the filamentary network around the Fornax cluster. In total, 81\% of galaxies from our subset belong to filaments and 19\% of galaxies are located outside filaments. We obtained the groups and clusters from the group catalogues by \citet{Makarov2011} and \citet{Kourkchi2017} and indicate those that are located in our dataset, in the lower panel of Fig.~\ref{Filaments_fig}. Apart from groups and clusters within filaments, we find that $\sim$30\% of galaxies are isolated in filaments, that is, they are in \textit{pristine filamentary} environments. With regard to the filaments around the Fornax-Eridanus Complex (486 galaxies), 33\% of them are galaxies in \textit{pristine filamentary} environments. 

We estimated the length of the central axis of each filament, that is the embedded graph from its 3D Cartesian plane, and present these estimates in Table \ref{tab:my_table1}. The length of the filament-axes vary from 4~--~12~Mpc/$h$. The longest of them are the axis of Filament~12 $\sim 12.41$~Mpc/$h$ comprising 14 galaxies and Filament~25 $\sim 10.35$~Mpc/$h$ with 41 galaxies. We define filaments shorter than 5 Mpc/$h$ as ``minor filaments''.

\subsection{Known groups and clusters in filaments} \label{sec:major_groups}
The large-scale structure around the Fornax cluster amasses various filamentary environments.  These filamentary environments are heterogeneous in nature, comprising galaxy groups, clusters, and pristine galaxies. Here, we detail this heterogeneity. We find that twenty out of twenty-seven filaments contain galaxy pairs, triplets, groups, and clusters; we list them in Table \ref{tab:my_table1}. As a reminder, we consider group and cluster galaxies as bound galaxies inside one virial radius (T16a). The most massive among these systems is the Fornax cluster (NGC~1399), located in Filament 15 and Filament~19. The latter filament connects the main Fornax cluster to its subgroup Fornax~A (NGC~1316). The NGC~1532 group is another system supposedly merging with the Fornax cluster and is part of Filament~19. The second most massive structure is the Eridanus supergroup (NGC~1302, NGC~1332,  NGC~1395, NGC~1398, NGC~1407) and is part of two filaments, Filament~18 and Filament~15, extending from the Fornax cluster. The Dorado group (NGC~1553, NGC~1672) is connected to the Fornax cluster through Filament~23 and Filament~25. \par
An extension of the Southern Supercluster is the Cetus-Aries Cloud (NGC~0584, NGC~0681, NGC~0720, NGC~0908, NGC~0936, NGC~1068; see \citealt{Tully1987a}) and is connected to the Fornax-Eridanus Complex through Filament~22. Seven minor ($l <$~5~Mpc/$h$ ) filaments connect groups within the Cetus-Aries Cloud and some of these groups are located at intersecting nodes (NGC~0720, NGC~0936, NGC~1068). The Cetus I (NGC~1068) Cloud is part of Filament~16 while Cetus II (NGC~0584, NGC~0681) Cloud is part of Filament~1. Among these minor filaments, Filament~12 encompasses only one bright galaxy, and connects the Cetus II cloud to the NGC~0720 group through Filament~8 and to NGC~0908 group through Filament~22. 

Known galaxy-groups within a distance of 10~Mpc from the Milky Way are NGC~0672, NGC~3031, NGC~5457 \citep[see][]{devauc75}. The latter two are part of Filament~10, while NGC~0672 is part of Filament~0. These filaments are connected to Fornax and Dorado Clouds, through a series of filaments comprising mostly pristine galaxies and some galaxy-pairs; they are Filament~3, Filament~5, Filament~24. \par 
Except for the aforementioned groups and clusters in the Fornax-Eridanus Complex, Cetus-Aries Clouds, and M31 (NGC~3031) group,  all other systems are pairs, triplets, and smaller groupings ($N_{gal} \leq 5 $) of  galaxies. Eleven filaments that we detect including the longest, that is, Filament~9, comprise only pristine galaxies  (Filament~5, Filament~12, Filament~13, Filament~22, Filament~26), a few pairs (Filament~4, Filament~9, Filament~14, Filament~21), and triplets (Filament~4, Filament~23)\footnote{Table~\ref{tab:my_table1} shows only the known groups/clusters from the group catalogues by \citet{Makarov2011} and \citet{Kourkchi2017}. However, some filaments contain more groups. For example, based on the catalogue by T16a, Filament~3 contains a group with $N_{gal} = 5$, but is not tabulated here.}. On the whole, the large-scale structure around the Fornax cluster we detected encompasses various filamentary environments.
\begin{figure}
\centering
   \includegraphics[scale=0.57]{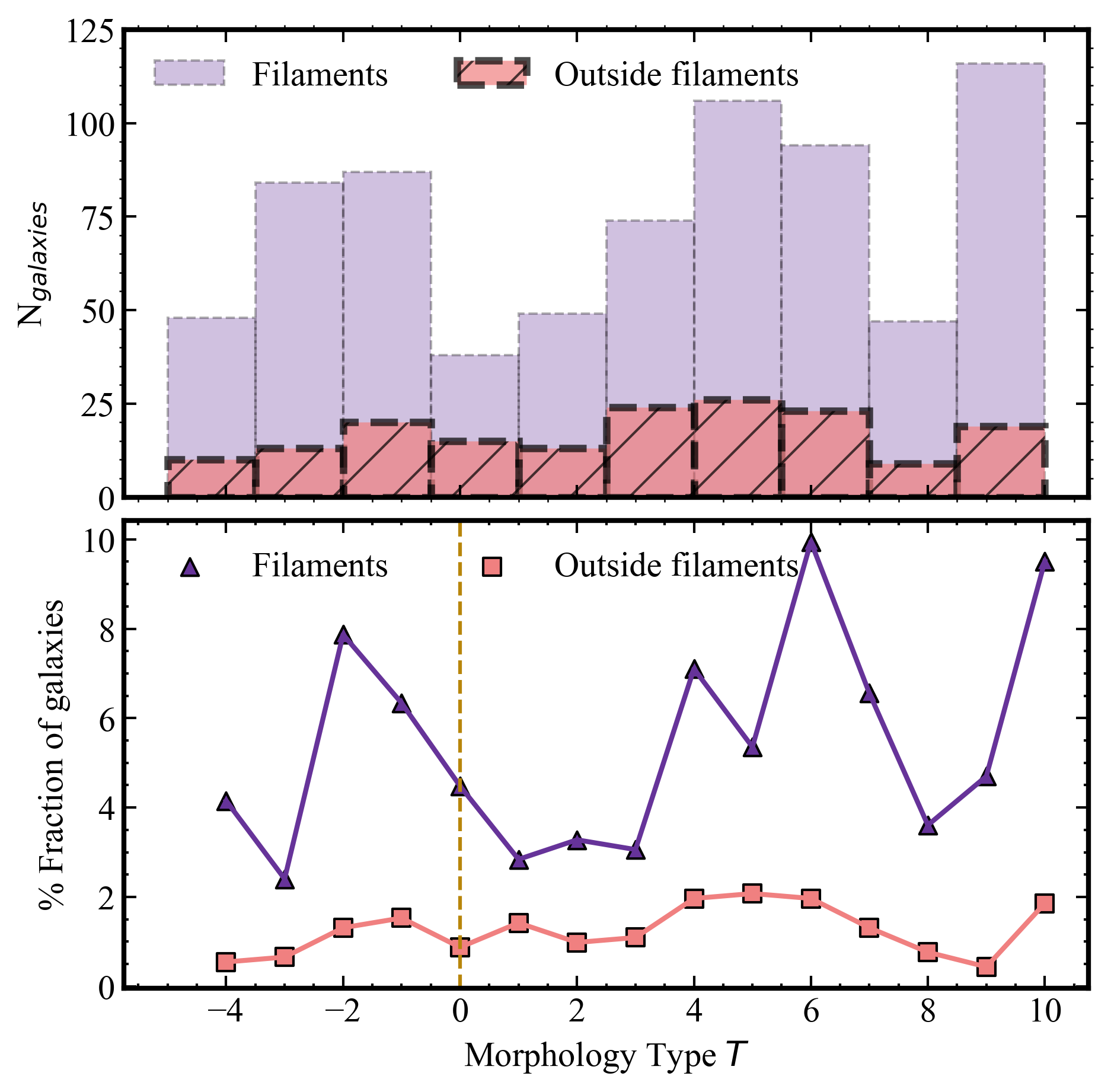}
      \caption{Histogram of galaxy-morphology in filaments and outside filaments. \textit{Top panel}: Purple indicates galaxies within filaments and coral dashed histogram indicates galaxies outside filaments. \textit{Lower panel}: Purple triangles represent galaxies in filaments and coral squares represent those outside filaments. Yellow dashed line at $T=0$ indicates the separation between ETGs and LTGs. } 
         \label{Morph_out}
\end{figure}
\subsection{The Fornax Wall}\label{FW sect}
The Fornax Wall was first discovered by \citet{Fairall94} and the Fornax-Eridanus supercluster is embedded within it. \citet{Fairall94} described \textit{Great Walls} as a large concentration of galaxies in a volume of several thousand km~s$^{-1}$ in redshift space. 

We identified the central axis of this structure that appeared six out of nine iterations with MMCrawling. The four filamentary segments that together constitute this topological feature are Filament~15, Filament~18, Filament~23, and Filament~25.   We show two perspectives of the Fornax Wall in Fig.~\ref{FS} and indicate all the major groups and clusters (Dorado group, Fornax cluster and its subgroup -- Fornax~A, Eridanus supergroup) in this region. We also indicate Filament~19 (length $\sim$~6~Mpc/$h$) which connects the Fornax cluster to its subgroup Fornax~A. We estimated the length of the central axis of this structure to be 31~Mpc/$h$ and it hosts 216 galaxies with log$_{10} (M_{*}/M_{\odot}) > 8.45$ within a neighbourhood radius of 3~Mpc/$h$. 

\begin{table*}[h]
\caption{Galaxy population in cosmic environments}      \label{tab:gal_cosm}     
\centering                                      
\begin{tabular}{ccccccc}        
\hline\hline    
Morphology  & & Filaments & & & Outside filaments &   \\  
 &       &   \% fraction     &           &       &      \% fraction        & \\
 \hline   
& Groups &  Pristine &  galaxy-pairs &  Groups &  Isolated & galaxy-pairs \\
 
(1) & & (2) & & & (3) &  \\
 \hline   
   \\
ETG &      15 &       7 &          4 &     -- &      5 &         $< 1$ \\
LTG &      23 &       23 &          9&     -- &      12 &        2 \\
\hline
\hline
\\ 
\end{tabular}
\tablefoot{Column 1 -- Morphology class; Column  2 -- Percentage fraction of galaxies in filaments, of which belong to groups/clusters, pristine environments, and galaxy-pairs; Column 3 --  Percentage fraction of galaxies outside filaments, of which belong to groups/clusters, isolated environments, and galaxy-pairs. }
\end{table*}
With 1-DREAM, the Fornax Wall was detected as a 1-dimensional feature and not as a 2-dimensional one as intuitively expected for cosmic walls. While the literature has identified this structure to be a cosmic wall, further analysis is required to comprehend the nature of it. We speculate that the structure we identified could be a filament embedded in a wall. In a forthcoming article (Raj et al. in prep), we will discuss this in detail.
\begin{figure}
\centering
   \includegraphics[scale=0.4]{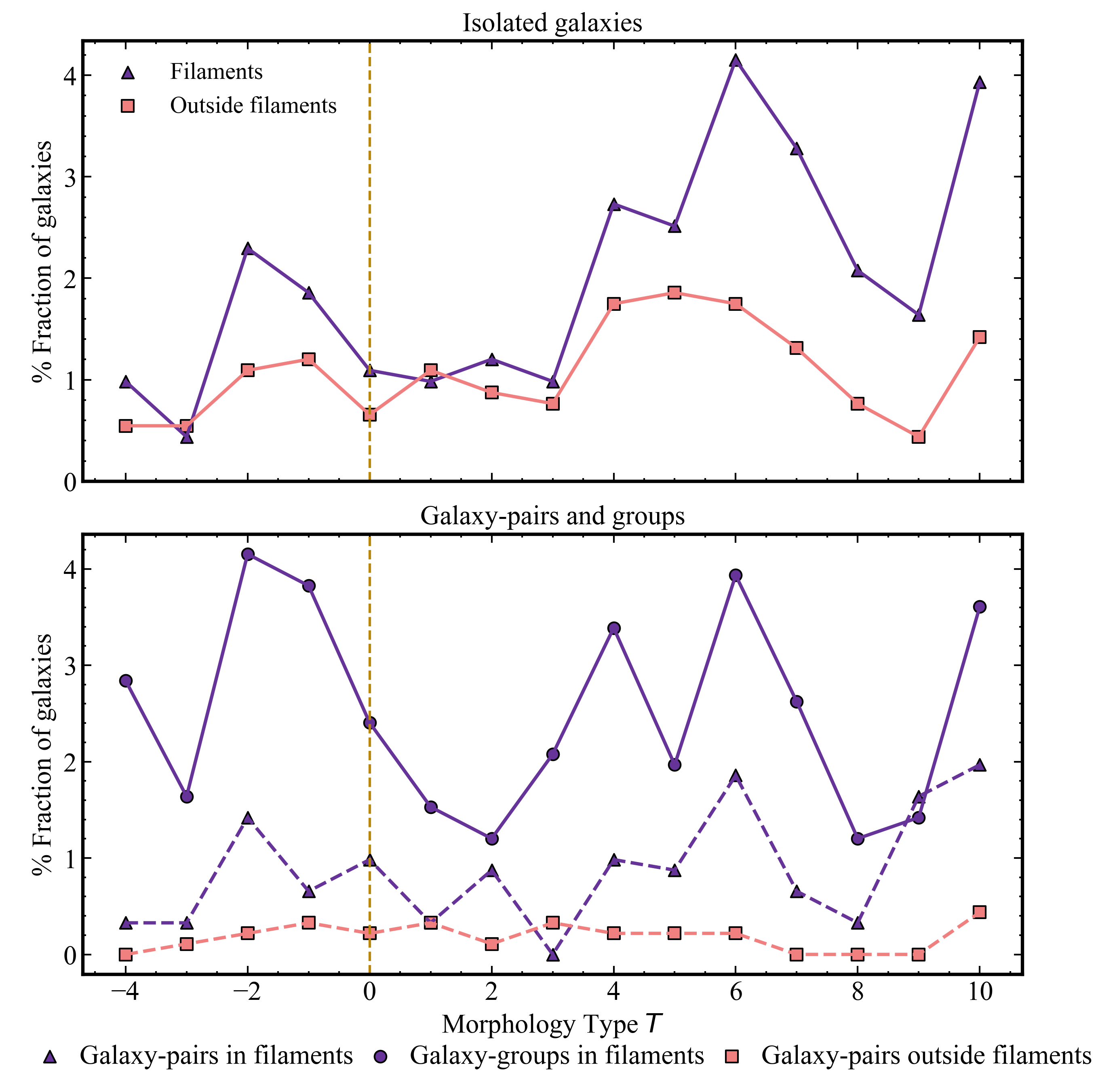}
      \caption{Histogram of galaxy-morphology in different environments. In both panels, we show the percentage fraction of $T$-Type  with regard to the total galaxy population, and we indicate the separation between ETGs and LTGs with a yellow dashed line, that is, at $T=0$; \textit{Top panel}: purple triangles indicate galaxies in pristine filamentary environments and coral squares indicate those outside filaments; \textit{Lower panel}: purple triangles (\textit{solid lines}) indicate galaxy-pairs in filaments, purple circles (\textit{dashed lines}) indicate galaxy-groups in filaments, and coral squares indicate galaxy-pairs outside filaments.}  
         \label{Morph_out_env}
\end{figure}

\section{Morphology-density relation of galaxies in filaments} \label{sec:Morphology}
In this section, we investigate the morphology-density relation of galaxies in the large-scale structure around the Fornax-Eridanus Complex with particular emphasis on galaxies part of the Fornax Wall. To do this, we segregate the catalogue into two main classes of galaxy morphology, defined by their Morphology Type $T$ (obtained from \citealt{Kourkchi2017} \footnote{We do not have $T$-types for 17 galaxies in our subset, which accounts for only $\sim$2\% of the extracted subset.}); they are early-type galaxies (ETGs, $-6 \leq T \leq 0$) and late-type galaxies (LTGs, $ T > 0 $ ). Of the extracted subset, 69\% are LTGs and 31\% are ETGs. 
\begin{figure*}[h]
\sidecaption
   \includegraphics[width=12cm]{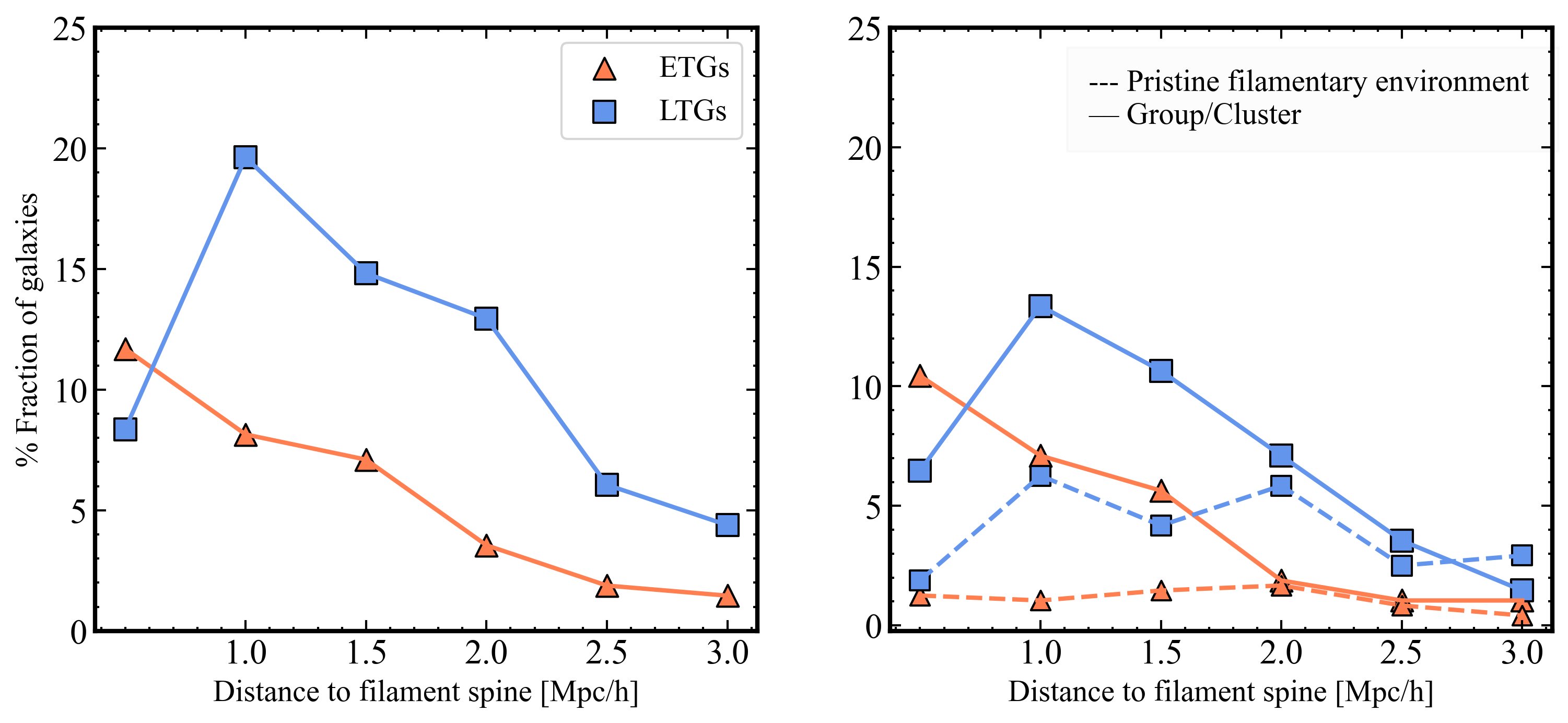}
      \caption{Morphology of galaxies in filaments around the Fornax-Eridanus Complex. In both panels, we show a histogram of ETGs (\textit{orange triangles}) and LTGs (\textit{blue squares}) as a function of their distance to the filament spine. In the \textit{right} panel, the dataset is divided in two categories, pristine filamentary galaxies (\textit{dashed lines}) and those associated to groups/clusters (\textit{solid line}).}  
         \label{Morph}
\end{figure*}
\begin{figure*}
\centering
   \includegraphics[width=\textwidth]{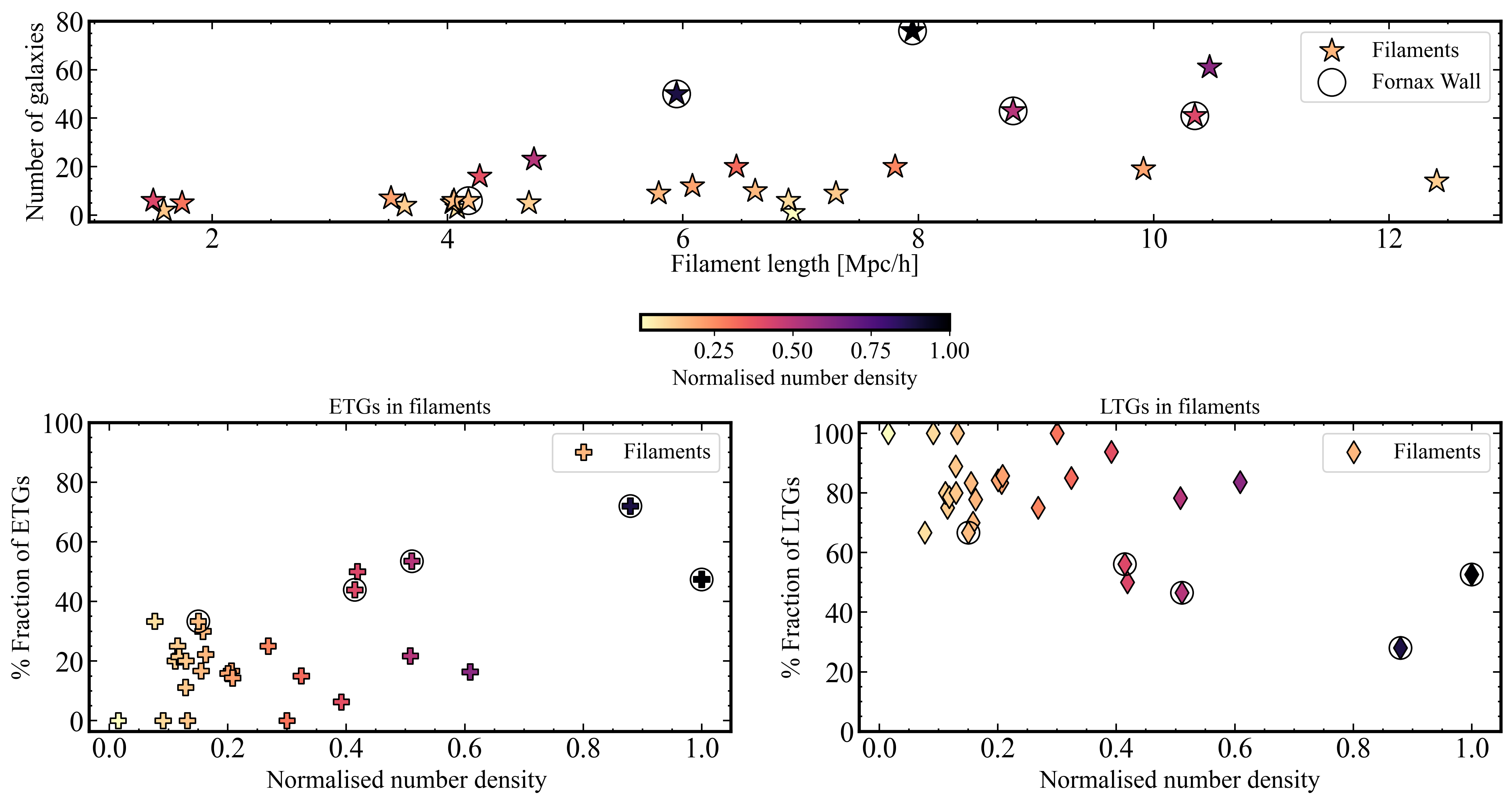}
      \caption{Morphology-density relation of galaxies in filaments around the Fornax-Eridanus Complex. \textit{Top panel}: Length of each filament (\textit{stars}) versus the number of galaxies belonging to it. For all panels, the colour bar indicates the normalised galaxy number-density of each filament ($N_{gal}/ l$), and \textit{circles} highlight filaments that constitute the Fornax Wall (includes Filament~19); \textit{Lower left panel}: Normalised galaxy number-density versus the percentage fraction of ETGs in each filament, represented by crosses; \textit{Lower right panel}: Same as the\textit{ lower left panel}, but for LTGs. Here, each diamond represents a filament.}      
         \label{Morph_fils}
\end{figure*}
\begin{figure*}[h]
\sidecaption
   \includegraphics[width=12cm]{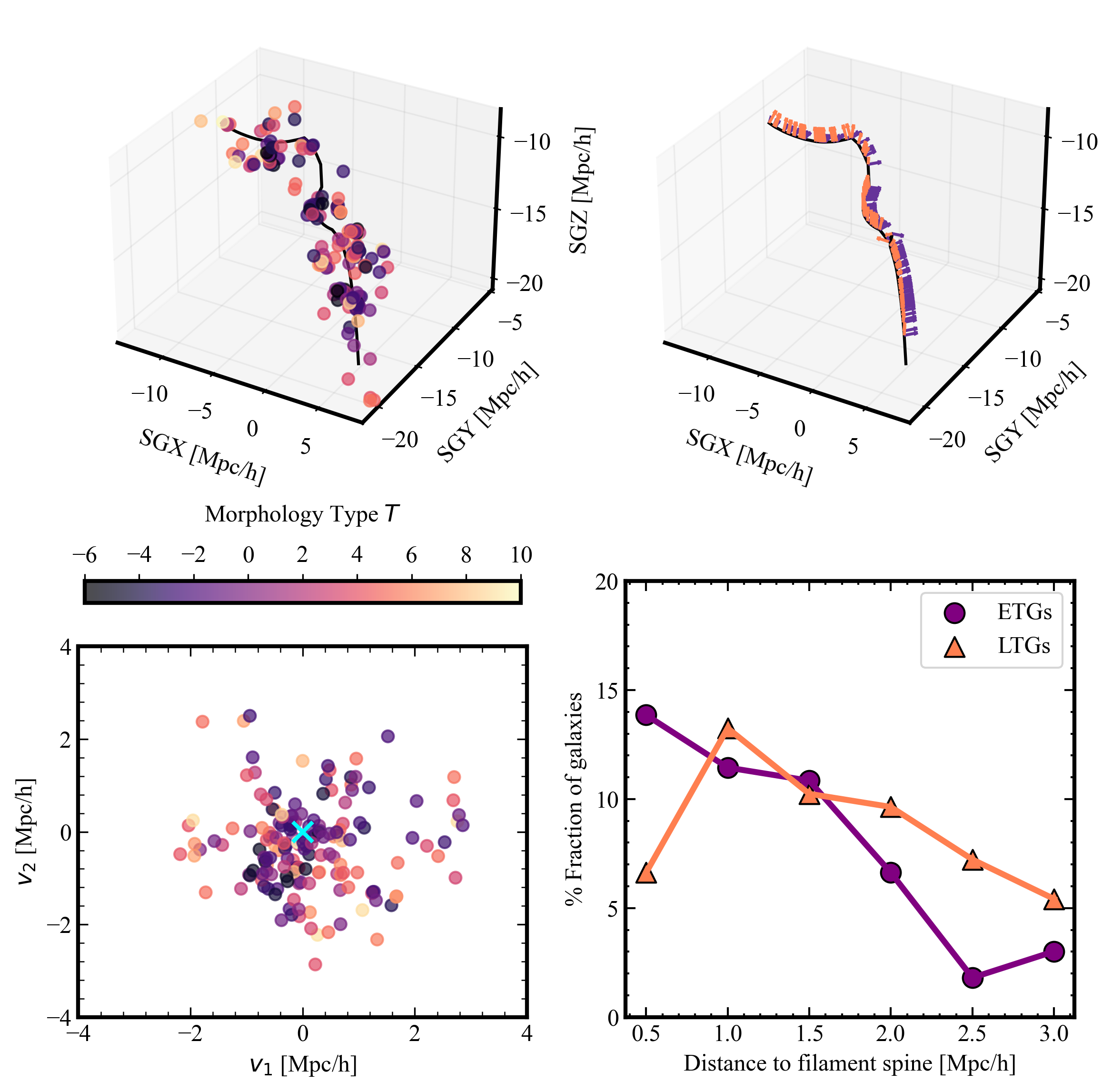}
      \caption{Morphology of galaxies and cross-sectional view of the Fornax Wall. \textit{Top left panel}: 3D view of the Fornax Wall in supergalactic coordinates. Each galaxy is highlighted by its Morphology Type $T$, as shown on the colour bar; \textit{Top right panel}: Normalised vectors for every galaxy on the central axis of the Fornax Wall in coral and purple; \textit{Lower left panel}: Cross-sectional view of all galaxies belonging to the Fornax Wall within a distance of 3~Mpc/$h$ to its centre. The centre is marked with a cyan cross and the galaxies are denoted by their Morphology Type $T$, as shown in the colour bar. \textit{Lower right panel}: Histogram of the ETGs (\textit{purple circles}) and LTGs (\textit{orange triangles}) as a function of their distance to the Fornax Wall.}
           
         \label{Morph_Fw}
   \end{figure*}
First, we investigate the effect of the cosmic environment on galaxy-morphology; the two cosmic environments are \textit{filaments} and \textit{outside filaments} (beyond 3~Mpc/$h$ of the filament spine). We show this in Fig.~\ref{Morph_out} and tabulate them in Table~\ref{tab:gal_cosm}.  In the top panel of Fig.~\ref{Morph_out}, we present the numbers of galaxies within each Morphology Type T bin for both cosmic environments (filaments and outside them). We also present the percentage fraction for the same in the lower panel, where we normalise our counts such that the sum of all points within this panel equals $100\%$. As mentioned in Sect.~\ref{Filaments FS}, $\sim$ 19\% of galaxies in our subset are located outside filaments. Among the galaxies outside filaments,  27\% of them are ETGs and 73\% are LTGs.  In Fig.~\ref{Morph_out_env}, we also indicate the environments within which these galaxies are located (\textit{isolated}, \textit{pairs}, \textit{groups}). Similarly, all percentages in each panel of this Figure sum to 100\%.  The fraction of ETGs outside filaments are mostly isolated (top panel of Fig.~\ref{Morph_out_env}) and $ < 1\%$ of the subset is ETGs in galaxy-pairs (lower panel of Fig.~\ref{Morph_out_env}). 

We do not find any groups ($N_{gal} > 2$) outside filaments. On the other hand, the fraction of ETGs is the highest in groups/clusters (15\%) within filaments. Concerning LTGs in filaments, we find that their fraction (with regard to the subset) is the same in pristine filamentary environments (top panel of Fig.~\ref{Morph_out_env}) and groups/clusters (lower panel of Fig.~\ref{Morph_out_env}), which is $\sim$ 23\% . We speculate that the galaxies outside filaments could belong to voids and a detailed study of this will be presented in a forthcoming article (Raj et al. in prep). From the literature \citep[e.g.][]{Cautun14, Awad2023}, the number density of galaxies in voids is lowest among other cosmic environments.

Now, we examine the role of galaxies' proximity to the filament core on their morphology. We determine the fraction of each morphology class with respect to the entire population of filament galaxies (from Table~\ref{tab:my_table1}), in bins of 0.5~Mpc/h from the filament core (we tabulate the number of galaxies in their respective bins in Table~\ref{tab:gal_pop}). In Fig.~\ref{Morph} (left panel), we show that the fraction of the ETG-population decreases with distance to the filament and 27\% of filament galaxies are ETGs in groups/clusters (see right panel of Fig.~\ref{Morph}). The fraction of the LTG-population ($\sim 8\%$) is lower than that of the ETG-population ($\sim 12\%$) at 0.5~Mpc from the filament spine. However, at 1~Mpc/$h$ from the filament spine, the LTG-population increases to 20\% (see left panel of Fig.~\ref{Morph}). Upon further inspection of this increment (at 1~Mpc), 13\% of these LTGs are in groups/clusters and 7\% are in pristine filamentary environments. We attribute this increment to LTGs which could be in outskirts of groups/clusters (see right panel of Fig.~\ref{Morph}) or to triplets/pairs. Of the total filament-population around the Fornax-Eridanus Complex, we find that 7\% are ETGs and 24\% are LTGs in pristine filamentary environments (see right panel of Fig.~\ref{Morph}), 27\% are ETGs are 42\% are LTGs in groups/clusters within filaments (refer to Table~\ref{tab:gal_pop}).
\begin{figure*}[h]
\sidecaption
   \includegraphics[width=12cm]{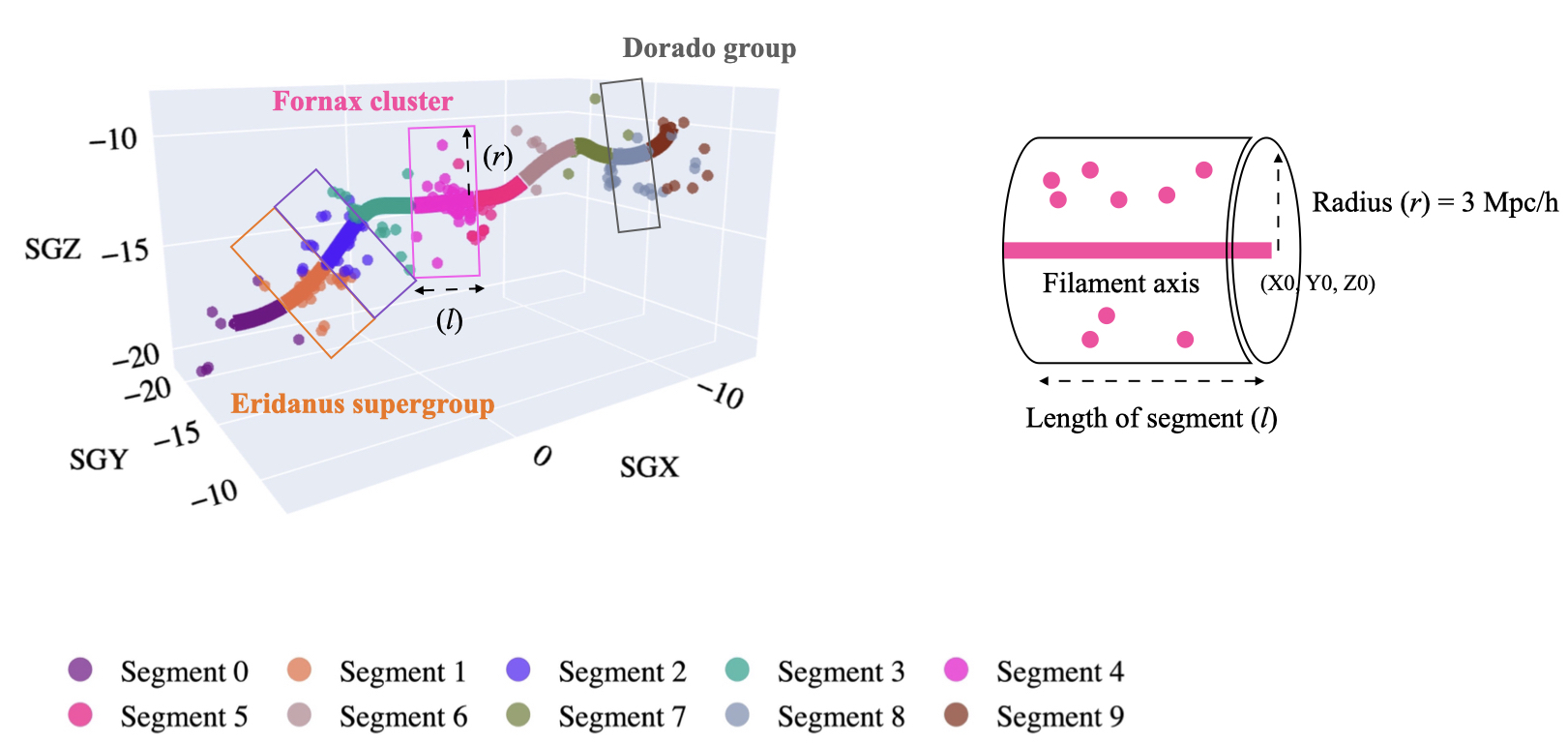}
      \caption{Segments of the Fornax Wall. \textit{Left~panel}: Distribution of galaxies part of the Fornax Wall, represented in supergalactic coordinates (SGX, SGY, SGZ). The Fornax Wall is divided into ten segments and each segment is highlighted with a different colour. The windows indicate the location of the major systems -- Eridanus supergroup,  Fornax cluster, and Dorado group in their respective segments. \textit{Right~panel}: Illustration of the method for estimating the length, radius, and volume of a segment's cylinder.}
           
         \label{FW_seg}
   \end{figure*}
 \begin{figure*}[h]
\centering
   \includegraphics[scale=0.48]{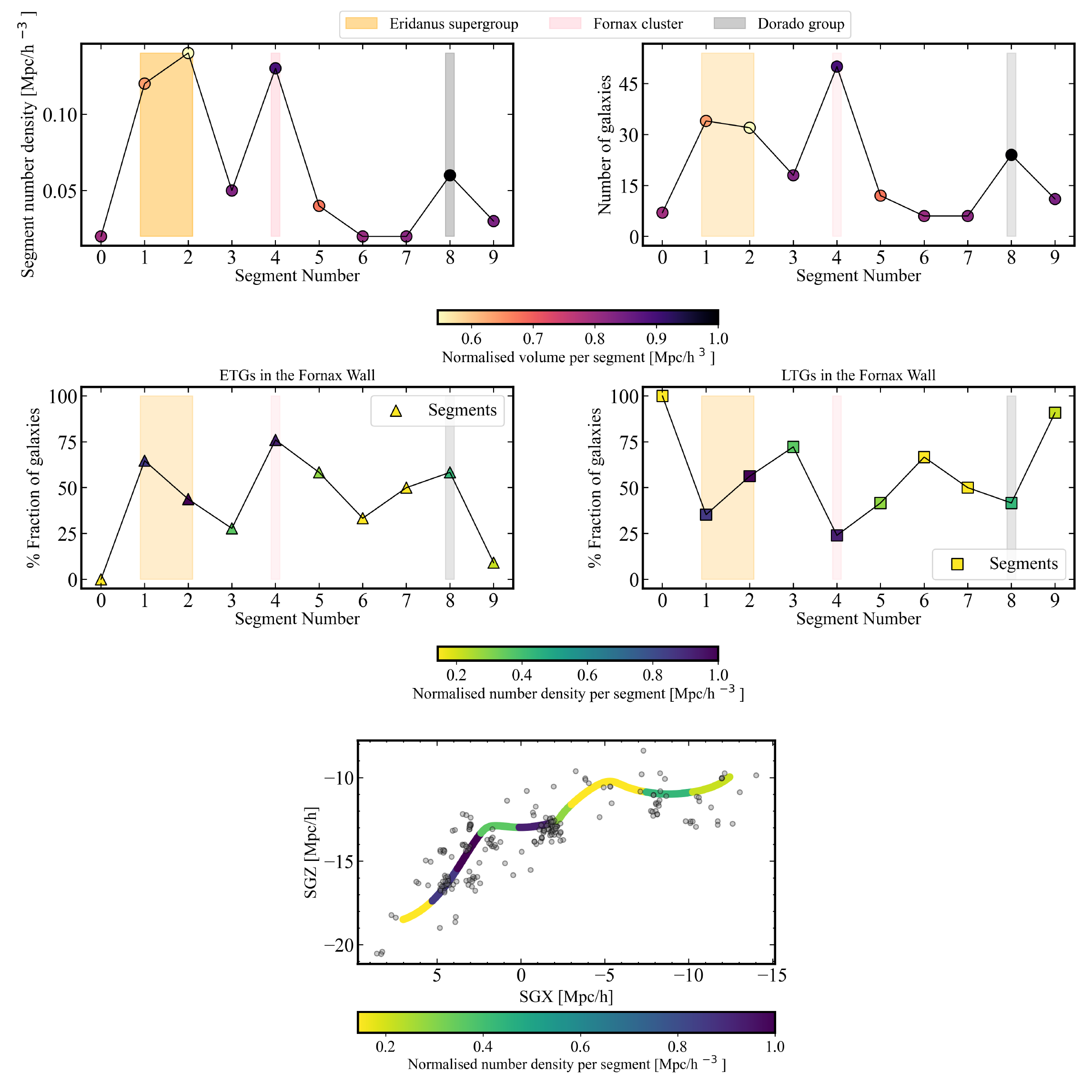}
      \caption{Morphology-density relation of galaxies part of the Fornax Wall. \textit{Top panel}: In both panels, $x$-axes represent the segment number corresponding to Fig.~\ref{FW_seg} and $y$-axis represents the segment number density on the \textit{left} and number of galaxies on the \textit{right} panel. The colour bar indicates the normalised volume per segment. The major systems as indicated in Fig.~\ref{FW_seg} are also highlighted here-- Eridanus supergroup in orange, Fornax cluster in pink, and Dorado group in grey; \textit{Middle panel}: Fraction of ETGs (\textit{left}) and LTGs (\textit{right}) in each segment. The colour bar indicates the normalised number density per segment; \textit{Lower panel}: The Fornax Wall and its constituent galaxies, projected in the supergalactic $x$-$z$ plane. Each segment is coloured according to its density; the colour bar depicts the normalised density.}
           
         \label{FW_ana}
   \end{figure*}
We also investigate the role of the filament number-density on the morphology of galaxies. In Fig.~\ref{Morph_fils} (top panel), we indicate that longer filaments have more galaxies than their shorter counterparts. From this, we estimate the number-density of each filament and indicate this with a colour bar in Fig.~\ref{Morph_fils} (top panel). The densest filaments include Filament~15, Filament~18, Filament~23, Filament~25, which together form the central axis of the Fornax Wall. In Fig.~\ref{Morph_fils}, we find that the fraction of ETGs (lower left panel) increases with the density of their hosting filament, while that of LTGs (lower right panel) decreases with increasing density of their filaments. This correlation indicates the effect of the local environment on morphology of galaxies, as proposed by \citet{dressler1980}.

Now we explain the effect of the large-scale environment on the morphology of galaxies part of the Fornax Wall. In the top-left panel of Fig.~\ref{Morph_Fw}, we show the 3D projection of the central axis of the Fornax Wall and indicate the Morphological Type $T$ of each galaxy in the colour bar. We also implemented a cross-sectional view along the Fornax Wall to inspect the distribution of galaxies surrounding its centre. The procedure explained below was developed by \citep{Canducci2022}  and is briefly recounted here. With this aim, we constructed a series of ortho-normal coordinate frames centred on the spine of the Wall and co-moving along its length. For every segment $\hat{\xi}_i = \frac{1}{|\xi_i|} (\xi^1_i, \xi^2_i, \xi^3_i)$ connecting two points that sample the Fornax Wall's central axis, we recovered two vectors $\vec{v}_1 = (v^1_1, v^2_1, v^3_1)$ and $\vec{v}_2 = (v^1_2, v^2_2, v^3_2)$ that are orthogonal to  $\hat{\xi_i}$ and that span the perpendicular plane to the axis of the Fornax Wall at the given point $i$. For finding these vectors, the following system of linear equations is solved:
\begin{equation}
\begin{cases} 
\vec{v_1} \cdot \vec{v_2} = v^1_1 v^1_2 + v^2_1 v^2_2 + v^3_1 v^3_2\\
\vec{v_1} \cdot \hat{\xi_i} =  v^1_1 \hat{\xi}^1_i + v^2_1 \hat{\xi}^2_i + v^3_1 \hat{\xi}^3_i\\
\vec{v_2} \cdot \hat{\xi_i} = v^1_2 \hat{\xi}^1_i + v^2_2 \hat{\xi}^2_i + v^3_2 \hat{\xi}^3_i.
\end{cases}
\end{equation}

This system of equations is degenerate and therefore permits an infinite number of solutions. Therefore, to keep a consistent orientation along the length of the central axis, we fixed the solution to be of the form:

\begin{equation}
    \vec{v}_1 = (\hat{\xi}^2_i, -\hat{\xi}^1_i, 0), \\ \vec{v}_2 = (\hat{\xi}^1_i \hat{\xi}^3_i, \hat{\xi}^2_i \hat{\xi}^3_i, -(\hat{\xi}^1_i)^2 -(\hat{\xi}^2_i)^2). 
\end{equation}

 The retrieved vectors are then normalised and are shown for every point on the central axis (top right panel of Fig.~\ref{Morph_Fw}). As mentioned, these vectors span the orthogonal plane to the axis at each given point $i$. Finally, we project the positions of the galaxies belonging to the Fornax Wall onto the orthogonal plane centred on the nearest point on the axis to each galaxy. These projected positions are then stacked to achieve the final cross-sectional view of all galaxies belonging to the Fornax Wall within a distance of 3~Mpc/$h$ to its centre. We show this result in the lower left panel of Fig.~\ref{Morph_Fw}, where the galaxies are also coloured by their Morphology Type $T$.
 
Similar to our results on the morphology-density relation of galaxies in filaments, we also see a morphological segregation of galaxies in the large-scale environment of the Fornax Wall. We find that $\sim$ 44\% (216) of galaxies which belong to the filamentary network of our data subset (486 galaxies) are associated to the Fornax Wall. In the lower-right panel of Fig.~\ref{Morph_Fw}, we see that the fraction of the LTG-population (7\%) is lower than that of ETGs (13\%) near the central axis of the filament $\sim$~0.5~Mpc/$h$. However, at 1~Mpc, the fraction of LTGs increases to 13\% and at 1.5~Mpc, the fraction of LTGs is similar to that of ETGs (11\%). Beyond this radius, the fraction of LTGs is higher than of that ETGs. 

From Fig.~\ref{Morph_Fw}, we can also see that the galaxy-number density of the Fornax Wall is larger within its centre (point $(0,0)$ in the lower left panel) and decreases gradually as we move radially away from it. This drop in the density of cosmic filaments as we move radially away from their centres has been shown previously in the form of radial density profiles of simulated cosmic filaments \citep{Ramsoy2021, Pfeifer2022, espinosa23, Awad2023}. Therefore, we observe the same correlation between density and distance away from the centre of the Fornax Wall.

\cite{Kuchner2022} showed that filament environments can vary in number-density and type (pristine or groups/cluster). Likewise, the galaxy-number density could vary across galaxy walls, owing to some of them within group/cluster or pristine environments. This number-density would imply different local environments across a filament/wall. Here, we explore the different environments across the Fornax Wall and its impact on its member galaxies. We do this by dividing the Wall into 10 segments (length 2.4--3.8~Mpc/$h$) and then estimate the volume and density of each segment. In Fig.~\ref{FW_seg}, we illustrate the segments and indicate the major clusters within the the Fornax Wall. 

We show the results of this analysis in Fig.~\ref{FW_ana}. Here, we describe each panel of this figure. In the top panel, we plot the number density of each segment (left) and their population number (right),  and find that the density peaks in segments containing the Eridanus supergroup, Fornax cluster, and the Dorado group. In the middle panel, we plot the percentage fraction of ETGs (left) and LTGs (right), and find that ETGs populate the densest regions of the Fornax Wall. Within the Eridanus supergroup, the NGC~1407 group is the most dynamically evolved (located in segment 1; \citealt{Brough06}) and 75\% of its galaxies are early-type and 25 \% are late-type; this is similar to the Fornax cluster. On the contrary, the region of this supergroup which comprises NGC~1395 (segment 2) has more (60\%) LTGs than ETGs. Lastly, the Dorado group has 60\% ETGs and 40\% LTGs. To support the argument that ETGs populate the densest regions of the Fornax Wall, we illustrate the galaxy number density in the lower panel of Fig.~\ref{FW_ana}. 

\begin{figure*} [h]
   \includegraphics[width=\textwidth]{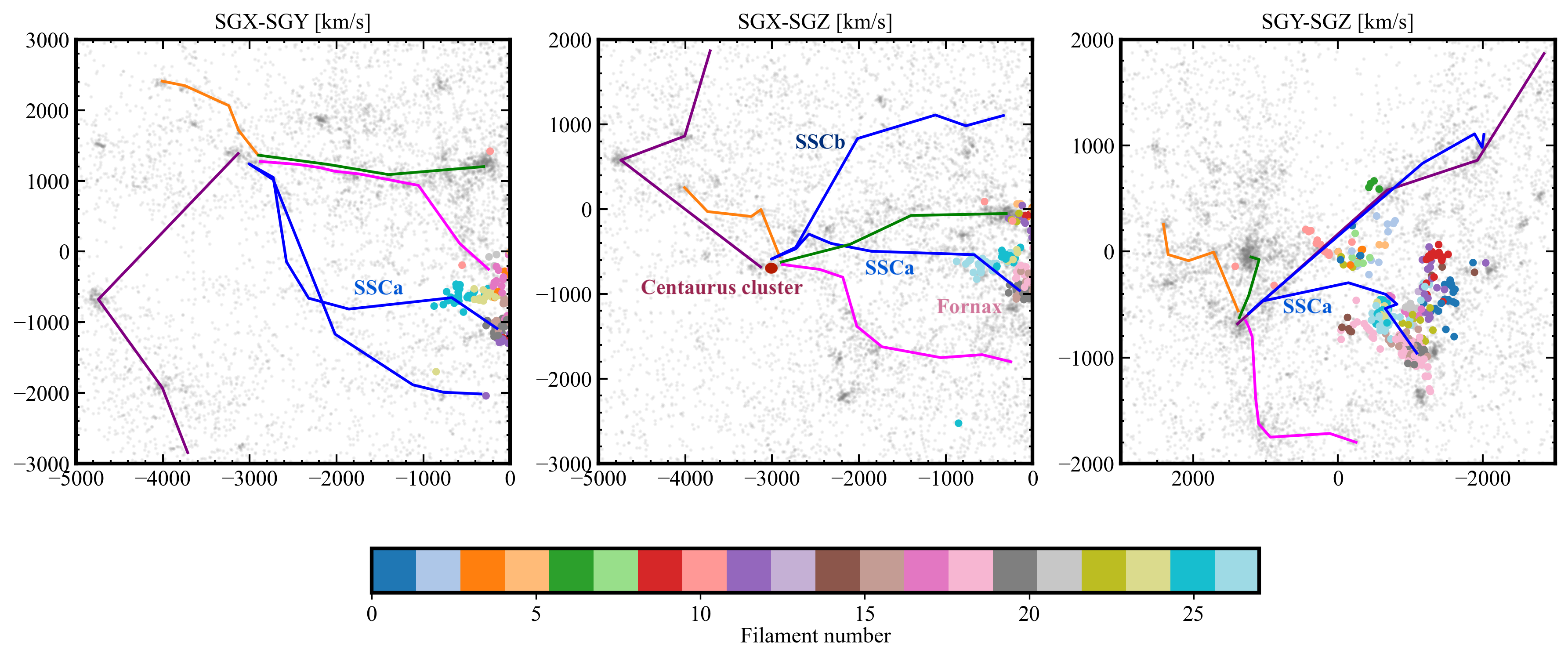}
      \caption{Cosmic web around the Fornax-Eridanus Complex. The V8k catalogue is over-plotted here (\textit{grey dots}) to show the extension of our detected filaments. Galaxies part of our dataset are highlighted with different colours based on their filament-membership; this is shown in the colour bar. The five filaments spines feeding the Centaurs cluster, as depicted by \citet{Courtois13}, are plotted here. Among them, the two branches of the Southern Supercluster Strand-- SSCa and SSCb, the Centaurus cluster, and the Fornax cluster are also marked in the centre panel. Each panel is projected in supergalactic coordinates along $x$-$y$, $x$-$z$, and $y$-$z$ planes. }
           
         \label{Courtois_LSS}
\end{figure*}

\section{Large-scale environment around the Fornax-Eridanus Complex} \label{sect:disc}
In this section, we compare our results to previous findings thereby investigating the role of the large-scale environment surrounding the Fornax-Eridanus Complex on its constituent galaxies. With this, we also show the extent of the detected filaments in the Cosmic Web. 

Over-densities in \textit{Great Walls} have been attributed to cores within superclusters embedded in them \citep{einasto16}. Gravitational instability theory \citep{Zeldovich1970, Icke1973, WhiteSilk1979} is the main contributor of the flow of matter within the Universe. This flow occurs from lower-density regions such as voids to higher-density regions such as walls. In this context, it is worth noting the presence of any cosmic voids surrounding the Fornax Wall which could play a role in shaping its structure. 

Using the Nexus+ methodology, \citet{Cautun2014} demonstrated this mode of matter transfer within the Millennium Simulations \citep{springel05, Boylan-Kolchin2009} evolved from redshift $z=2$ till the present. They also showed that any transport of mass that defies this rule is an artefact of the method used for structure-identification and thus does not challenge the underlying cosmological evolution theories. \citet{Cautun2014} also emphasised the strong outflow that occurs from voids towards sheets which act as void boundaries. As indicated by \citet{JonesJones1980, PhillippsDavies1992, Drinkwater2000}, a large void extending around 40~Mpc exists immediately beyond the Fornax cluster. The close presence of this void therefore fortifies the claim of the existence of a cosmic structure at its boundary (i.e. the Fornax Wall). It also suggests that many of the galaxies within the Fornax Wall have been the result of outflow occurring between the two environments. It is possible to verify this claim by looking at the dynamical behaviour of the galaxies within the wall and comparing the direction with which they are moving with respect to the Fornax Void. We leave such explorations for future work. 

Our analysis reveals a significant morphological segregation of ETGs dominating dense environments such as dynamically evolved groups/clusters of filaments. In fact, the ETG population decreases with increasing distance to the filament spine. Moreover, groups and clusters in the high-density core of the Fornax-Eridanus Complex are more ETG-rich than other environments in filaments connecting to the Complex. There has been evidence of the effect of the large-scale environment on galaxy properties, such that the core of superclusters are populated with red galaxies, while blue galaxies are homogeneously distributed in the outskirts of superclusters \citep[see][]{haines06, meinasto11, Einasto18, einasto22}. A morphological-segregation in superclusters can also be attributed to its evolutionary history \citep{Einasto08}.  In the case of the Fornax-Eridanus Complex, the merging of groups within it, further suggests that pre-processing of galaxies before cluster-infall is the dominating factor for the observed morphological segregation. Moreover, filaments outside the Complex have lower fractions of ETGs, which are mostly located in groups near the central axis of filaments. 

Filaments are considered to be an intermediate-density region and follow the morphology-density relation \citep[e.g][]{kuutuma17, santiago20, Hoosain24}. However, we suspect that the local density might not be the only factor contributing to a galaxy's morphology. An example of this is Filament~10, where the M81 group is located. Although this filament ($l \sim~10.5$~Mpc/$h$) has 63 galaxies, both in groups and pristine environments, the fraction of LTGs (85\%) is higher than that of ETGs (15\%). We speculate that another possible contributing factor to a galaxy's morphology could be its proximity to a filament core.  As most of the galaxies in the M81 group are beyond 1.5~Mpc/$h$ from the central axis of the filament, the primordial gas supply from the Cosmic Web \citep[see][]{calvo19, winkel21} might not be completely shut down.  A combination of several internal and external processes (e.g. ram-pressure stripping; \citealt{gunn72}) in galaxy-clusters can eventually diminish the gas content in galaxies. 

The large-scale structure around the Fornax-Eridanus Complex constitutes a variety of filamentary environments, not only in terms of galaxy number-density, but also its evolutionary history. Previously, \citet{Kuchner2022} demonstrated the heterogeneity of filamentary environments such that the filaments can be pristine or dominated by groups/clusters. In order to examine the effect of the filament itself on galaxy properties, the groups/clusters within it have to be excluded. We elucidated this in Fig.~\ref{Morph} (right panel) and found $ \sim $ 7\% of the galaxy-population in filaments are ETGs in pristine environments. On the other hand, the fraction of the LTG-population ($ \sim $ 24\%) in pristine environments of filaments is higher than that of ETGs. Further exploration of the stellar mass and gas content of these galaxies is required to understand the formation channels of ETGs in pristine environments of filaments.

\citet{Courtois13} showed that five filaments emerge from the Centaurus cluster, with one of them connected to the Southern Supercluster Strand SSCa (Fornax-Eridanus-Dorado).  In order to visualise the extent of the emerging filament SSCa and its connection to the filaments we detect, we plot them in Fig.~\ref{Courtois_LSS} \footnote{We matched the subset of the T16a dataset to the V8k catalogue from EDD \citep{Tully09} and this resulted in 386 out of 486 galaxies.}. Filament~25 is part of the Fornax Wall, and an extension of this filament constitutes SSCa, that connects to the Centaurs cluster. While the Fornax Wall itself has no clear boundary, the density contrast from this structure to its nearby filaments would suggest a filamentary network of lower galaxy number-density, that continues all the way to the Centaurus cluster. The other branch of the Southern Supercluster Strand is SSCb, which is also one of the emerging filaments from the Centaurus cluster (see Fig.~\ref{Courtois_LSS}). This branch is connected to the Fornax-Eridanus Complex through the Cetus-Aries Cloud \citep{tully87}. Although, \citet{tully87} indicated that the Cetus-Aries Cloud is a minor filament, our results reveal that several small filaments connect groups within this Cloud, as explained in Sect.~\ref{sec:major_groups}.

\section{Summary and Conclusions} \label{sect: summary}
In this work, we explored the Fornax-Eridanus Complex using the novel ML toolbox, 1-DREAM. We then presented the analysis of the filamentary network that we detected within the region. This analysis included an investigation of the influence of filamentary environments on the morphology of galaxies, defined by the Morphological Type $T$ parameter. Finally, we discussed the large-scale environment around the Fornax-Eridanus Complex. We enumerate the main results of this study below:

\begin{enumerate}
    \item With 1-DREAM, we identified a filamentary network of galaxies around the Fornax-Eridanus Complex. Of the extracted volume, 81\% are galaxies located in filaments and 19\% are galaxies outside filaments. The filamentary network around the Fornax-Eridanus Complex comprises 27 filaments that vary in length and galaxy number-density. These filaments exhibit various environments, which are defined by their number-density and groups/clusters embedded in them. Among them, some contain only pristine filament galaxies. 

\item The subsequent analysis of the aforementioned filaments revealed a well-known structure-- the Fornax Wall, which encompasses the Fornax-Eridanus-Dorado supercluster. The length of the central axis of this Wall is $\sim$ 31~Mpc/$h$ and hosts 216 galaxies with log$_{10} (M_{*}/M_{\odot}) > 8.45$ within a neighbourhood radius of 3~Mpc/$h$.

\item The literature suggests that the morphology-density relation of galaxies extends to both filaments and walls. Our analysis corroborates this by showing a decrease in the fraction of the ETG-population ($-6 \leq T \leq 0$) with increasing distances from the filament spine. The fraction of the LTG-population ($T > 0$; 8\%) in our catalogue is lower than that of ETGs (12\%) at 0.5~Mpc from the filament spine. We attribute the increment (20\%) at 1~Mpc/$h$ to LTGs located in the outskirts of groups/clusters as well as galaxy-triplets/pairs. Of the total galaxy-population in filaments around the Fornax-Eridanus Complex, $\sim$ 27\% are ETGs and $\sim$ 42\% are LTGs in groups/clusters within filaments, while $\sim$ 7\% are ETGs and $\sim$ 24\% are LTGs located in pristine environments of filaments. Further investigation of the gas content and stellar mass of ETGs is required to understand their formation channels in pristine environments. 

\item We observed that 44\% of the galaxies from the filamentary network belong to the Fornax Wall. Our findings show that the galaxy number-density varies across the Fornax Wall and therefore impacts its member-galaxies differently. As such, ETGs dominate the densest regions of the Fornax Wall.

\end{enumerate}

In conclusion, this study has revealed the Cosmic Web around the Fornax cluster, displaying various filamentary environments within and around the supercluster. With this, our research reaffirms the heterogeneity of filamentary environments that was proposed by \citet{Kuchner2022}. When investigating the impact of the local environment on galaxy morphology, number-density as well as proximity to the filament core (central axis) are to be considered. In this context, the morphological segregation in the Fornax Wall can be attributed to pre-processing in groups embedded in it and the inflow of galaxies from the surrounding Fornax Void.  Lastly, we restate that 1-DREAM is a reliable ML tool to model the central axis of filaments, which in turn is essential to study the role of the environment on galaxy properties. As such, our future endeavours entail an in-depth study of the structure of the Fornax Wall as well as some observational properties of different galaxy populations pertaining to the filaments we analysed here.

\begin{acknowledgements}
The authors are grateful to the referee for their constructive feedback which helped to improve this article. This project is funded by the European Union’s Horizon 2020 research and innovation programme under the Marie Skłodowska-Curie grant agreement No 101066353 (ELATE). P.A. acknowledges support by the DSSC Doctoral Training Program of the University of Groningen. This work made use of Astropy: \url{http://www.astropy.org} a community-developed core Python package and an ecosystem of tools and resources for astronomy \citep{astropy:2013, astropy:2018, astropy:2022}. We used Matplotlib \citep{Hunter:2007} and Plotly \citep{plotly}: \url{https://plot.ly} to make interactive 3D plots. 
    
\end{acknowledgements}

\bibliographystyle{aa.bst}
 \bibliography{FS1} 

\begin{thebibliography}{133}
\expandafter\ifx\csname natexlab\endcsname\relax\def\natexlab#1{#1}\fi

\bibitem[{{Aghanim} {et~al.}(2024){Aghanim}, {Tuominen}, {Bonjean}, {Gouin},
  {Bonnaire}, \& {Einasto}}]{aghanim24}
{Aghanim}, N., {Tuominen}, T., {Bonjean}, V., {et~al.} 2024, arXiv e-prints,
  arXiv:2402.18455

\bibitem[{{Arag{\'o}n-Calvo} {et~al.}(2007){Arag{\'o}n-Calvo}, {Jones}, {van de
  Weygaert}, \& {van der Hulst}}]{calvo07}
{Arag{\'o}n-Calvo}, M.~A., {Jones}, B.~J.~T., {van de Weygaert}, R., \& {van
  der Hulst}, J.~M. 2007, \aap, 474, 315

\bibitem[{{Aragon Calvo} {et~al.}(2019){Aragon Calvo}, {Neyrinck}, \&
  {Silk}}]{calvo19}
{Aragon Calvo}, M.~A., {Neyrinck}, M.~C., \& {Silk}, J. 2019, The Open Journal
  of Astrophysics, 2, 7

\bibitem[{{Araya-Melo} {et~al.}(2009){Araya-Melo}, {Reisenegger}, {Meza}, {van
  de Weygaert}, {D{\"u}nner}, \& {Quintana}}]{araya09}
{Araya-Melo}, P.~A., {Reisenegger}, A., {Meza}, A., {et~al.} 2009, \mnras, 399,
  97

\bibitem[{{Astropy Collaboration} {et~al.}(2022){Astropy Collaboration},
  {Price-Whelan}, {Lim}, {Earl}, {Starkman}, {Bradley}, {Shupe}, {Patil},
  {Corrales}, {Brasseur}, {N{"o}the}, {Donath}, {Tollerud}, {Morris},
  {Ginsburg}, {Vaher}, {Weaver}, {Tocknell}, {Jamieson}, {van Kerkwijk},
  {Robitaille}, {Merry}, {Bachetti}, {G{"u}nther}, {Aldcroft},
  {Alvarado-Montes}, {Archibald}, {B{'o}di}, {Bapat}, {Barentsen}, {Baz{'a}n},
  {Biswas}, {Boquien}, {Burke}, {Cara}, {Cara}, {Conroy}, {Conseil}, {Craig},
  {Cross}, {Cruz}, {D'Eugenio}, {Dencheva}, {Devillepoix}, {Dietrich},
  {Eigenbrot}, {Erben}, {Ferreira}, {Foreman-Mackey}, {Fox}, {Freij}, {Garg},
  {Geda}, {Glattly}, {Gondhalekar}, {Gordon}, {Grant}, {Greenfield}, {Groener},
  {Guest}, {Gurovich}, {Handberg}, {Hart}, {Hatfield-Dodds}, {Homeier},
  {Hosseinzadeh}, {Jenness}, {Jones}, {Joseph}, {Kalmbach}, {Karamehmetoglu},
  {Ka{l}uszy{'n}ski}, {Kelley}, {Kern}, {Kerzendorf}, {Koch}, {Kulumani},
  {Lee}, {Ly}, {Ma}, {MacBride}, {Maljaars}, {Muna}, {Murphy}, {Norman},
  {O'Steen}, {Oman}, {Pacifici}, {Pascual}, {Pascual-Granado}, {Patil},
  {Perren}, {Pickering}, {Rastogi}, {Roulston}, {Ryan}, {Rykoff}, {Sabater},
  {Sakurikar}, {Salgado}, {Sanghi}, {Saunders}, {Savchenko}, {Schwardt},
  {Seifert-Eckert}, {Shih}, {Jain}, {Shukla}, {Sick}, {Simpson},
  {Singanamalla}, {Singer}, {Singhal}, {Sinha}, {Sip{H{o}}cz}, {Spitler},
  {Stansby}, {Streicher}, {{{S}}umak}, {Swinbank}, {Taranu}, {Tewary},
  {Tremblay}, {Val-Borro}, {Van Kooten}, {Vasovi{'c}}, {Verma}, {de Miranda
  Cardoso}, {Williams}, {Wilson}, {Winkel}, {Wood-Vasey}, {Xue}, {Yoachim},
  {Zhang}, {Zonca}, \& {Astropy Project Contributors}}]{astropy:2022}
{Astropy Collaboration}, {Price-Whelan}, A.~M., {Lim}, P.~L., {et~al.} 2022,
  \apj, 935, 167

\bibitem[{{Astropy Collaboration} {et~al.}(2018){Astropy Collaboration},
  {Price-Whelan}, {Sip{\H{o}}cz}, {G{\"u}nther}, {Lim}, {Crawford}, {Conseil},
  {Shupe}, {Craig}, {Dencheva}, {Ginsburg}, {Vand erPlas}, {Bradley},
  {P{\'e}rez-Su{\'a}rez}, {de Val-Borro}, {Aldcroft}, {Cruz}, {Robitaille},
  {Tollerud}, {Ardelean}, {Babej}, {Bach}, {Bachetti}, {Bakanov}, {Bamford},
  {Barentsen}, {Barmby}, {Baumbach}, {Berry}, {Biscani}, {Boquien}, {Bostroem},
  {Bouma}, {Brammer}, {Bray}, {Breytenbach}, {Buddelmeijer}, {Burke},
  {Calderone}, {Cano Rodr{\'\i}guez}, {Cara}, {Cardoso}, {Cheedella}, {Copin},
  {Corrales}, {Crichton}, {D'Avella}, {Deil}, {Depagne}, {Dietrich}, {Donath},
  {Droettboom}, {Earl}, {Erben}, {Fabbro}, {Ferreira}, {Finethy}, {Fox},
  {Garrison}, {Gibbons}, {Goldstein}, {Gommers}, {Greco}, {Greenfield},
  {Groener}, {Grollier}, {Hagen}, {Hirst}, {Homeier}, {Horton}, {Hosseinzadeh},
  {Hu}, {Hunkeler}, {Ivezi{\'c}}, {Jain}, {Jenness}, {Kanarek}, {Kendrew},
  {Kern}, {Kerzendorf}, {Khvalko}, {King}, {Kirkby}, {Kulkarni}, {Kumar},
  {Lee}, {Lenz}, {Littlefair}, {Ma}, {Macleod}, {Mastropietro}, {McCully},
  {Montagnac}, {Morris}, {Mueller}, {Mumford}, {Muna}, {Murphy}, {Nelson},
  {Nguyen}, {Ninan}, {N{\"o}the}, {Ogaz}, {Oh}, {Parejko}, {Parley}, {Pascual},
  {Patil}, {Patil}, {Plunkett}, {Prochaska}, {Rastogi}, {Reddy Janga},
  {Sabater}, {Sakurikar}, {Seifert}, {Sherbert}, {Sherwood-Taylor}, {Shih},
  {Sick}, {Silbiger}, {Singanamalla}, {Singer}, {Sladen}, {Sooley},
  {Sornarajah}, {Streicher}, {Teuben}, {Thomas}, {Tremblay}, {Turner},
  {Terr{\'o}n}, {van Kerkwijk}, {de la Vega}, {Watkins}, {Weaver}, {Whitmore},
  {Woillez}, {Zabalza}, \& {Astropy Contributors}}]{astropy:2018}
{Astropy Collaboration}, {Price-Whelan}, A.~M., {Sip{\H{o}}cz}, B.~M., {et~al.}
  2018, \aj, 156, 123

\bibitem[{{Astropy Collaboration} {et~al.}(2013){Astropy Collaboration},
  {Robitaille}, {Tollerud}, {Greenfield}, {Droettboom}, {Bray}, {Aldcroft},
  {Davis}, {Ginsburg}, {Price-Whelan}, {Kerzendorf}, {Conley}, {Crighton},
  {Barbary}, {Muna}, {Ferguson}, {Grollier}, {Parikh}, {Nair}, {Unther},
  {Deil}, {Woillez}, {Conseil}, {Kramer}, {Turner}, {Singer}, {Fox}, {Weaver},
  {Zabalza}, {Edwards}, {Azalee Bostroem}, {Burke}, {Casey}, {Crawford},
  {Dencheva}, {Ely}, {Jenness}, {Labrie}, {Lim}, {Pierfederici}, {Pontzen},
  {Ptak}, {Refsdal}, {Servillat}, \& {Streicher}}]{astropy:2013}
{Astropy Collaboration}, {Robitaille}, T.~P., {Tollerud}, E.~J., {et~al.} 2013,
  \aap, 558, A33

\bibitem[{{Awad} {et~al.}(2023){Awad}, {Peletier}, {Canducci}, {Smith},
  {Taghribi}, {Mohammadi}, {Shin}, {Ti{\v{n}}o}, \& {Bunte}}]{Awad2023}
{Awad}, P., {Peletier}, R., {Canducci}, M., {et~al.} 2023, \mnras, 520, 4517

\bibitem[{{Bag} {et~al.}(2023){Bag}, {Liivam{\"a}gi}, \& {Einasto}}]{Bag23}
{Bag}, S., {Liivam{\"a}gi}, L.~J., \& {Einasto}, M. 2023, \mnras, 521, 4712

\bibitem[{{Balogh} {et~al.}(2000){Balogh}, {Navarro}, \& {Morris}}]{Balogh2000}
{Balogh}, M.~L., {Navarro}, J.~F., \& {Morris}, S.~L. 2000, \apj, 540, 113

\bibitem[{{Bamford} {et~al.}(2009){Bamford}, {Nichol}, {Baldry}, {Land},
  {Lintott}, {Schawinski}, {Slosar}, {Szalay}, {Thomas}, {Torki}, {Andreescu},
  {Edmondson}, {Miller}, {Murray}, {Raddick}, \& {Vandenberg}}]{bamford09}
{Bamford}, S.~P., {Nichol}, R.~C., {Baldry}, I.~K., {et~al.} 2009, \mnras, 393,
  1324

\bibitem[{{Benavides} {et~al.}(2023){Benavides}, {Sales}, {Abadi}, {Marinacci},
  {Vogelsberger}, \& {Hernquist}}]{Benavides23}
{Benavides}, J.~A., {Sales}, L.~V., {Abadi}, M.~G., {et~al.} 2023, \mnras, 522,
  1033

\bibitem[{{Bidaran} {et~al.}(2022){Bidaran}, {La Barbera}, {Pasquali},
  {Peletier}, {van de Ven}, {Grebel}, {Falc{\'o}n-Barroso}, {Sybilska},
  {Gadotti}, \& {Coccato}}]{Bidaran22}
{Bidaran}, B., {La Barbera}, F., {Pasquali}, A., {et~al.} 2022, \mnras, 515,
  4622

\bibitem[{{Bond} {et~al.}(1996){Bond}, {Kofman}, \& {Pogosyan}}]{Bond1996}
{Bond}, J.~R., {Kofman}, L., \& {Pogosyan}, D. 1996, \nat, 380, 603

\bibitem[{{Bonnaire} {et~al.}(2020){Bonnaire}, {Aghanim}, {Decelle}, \&
  {Douspis}}]{Bonnaire20}
{Bonnaire}, T., {Aghanim}, N., {Decelle}, A., \& {Douspis}, M. 2020, \aap, 637,
  A18

\bibitem[{{Boselli} \& {Gavazzi}(2006)}]{Boselli06}
{Boselli}, A. \& {Gavazzi}, G. 2006, \pasp, 118, 517

\bibitem[{{Boylan-Kolchin} {et~al.}(2009){Boylan-Kolchin}, {Springel}, {White},
  {Jenkins}, \& {Lemson}}]{Boylan-Kolchin2009}
{Boylan-Kolchin}, M., {Springel}, V., {White}, S. D.~M., {Jenkins}, A., \&
  {Lemson}, G. 2009, \mnras, 398, 1150

\bibitem[{{Brough} {et~al.}(2006){Brough}, {Forbes}, {Kilborn}, {Couch}, \&
  {Colless}}]{Brough06}
{Brough}, S., {Forbes}, D.~A., {Kilborn}, V.~A., {Couch}, W., \& {Colless}, M.
  2006, \mnras, 369, 1351

\bibitem[{{Brown} {et~al.}(2023){Brown}, {Roberts}, {Thorp}, {Ellison},
  {Zabel}, {Wilson}, {Bah{\'e}}, {Bisaria}, {Bolatto}, {Boselli}, {Chung},
  {Cortese}, {Catinella}, {Davis}, {Jim{\'e}nez-Donaire}, {Lagos}, {Lee},
  {Parker}, {Smith}, {Spekkens}, {Stevens}, {Villanueva}, \& {Watts}}]{brown23}
{Brown}, T., {Roberts}, I.~D., {Thorp}, M., {et~al.} 2023, \apj, 956, 37

\bibitem[{{Bulichi} {et~al.}(2023){Bulichi}, {Dave}, \& {Kraljic}}]{bulichi23}
{Bulichi}, T.-E., {Dave}, R., \& {Kraljic}, K. 2023, arXiv e-prints,
  arXiv:2309.03282

\bibitem[{{Canducci} {et~al.}(2022){Canducci}, {Awad}, {Taghribi}, {Mohammadi},
  {Mastropietro}, {De Rijcke}, {Peletier}, {Smith}, {Bunte}, \&
  {Ti{\v{n}}o}}]{Canducci2022}
{Canducci}, M., {Awad}, P., {Taghribi}, A., {et~al.} 2022, Astronomy and
  Computing, 41, 100658

\bibitem[{Canducci {et~al.}(2022)Canducci, Tiño, \&
  Mastropietro}]{Canducci_2022_Prob}
Canducci, M., Tiño, P., \& Mastropietro, M. 2022, Artificial Intelligence,
  302, 103579

\bibitem[{{Castignani} {et~al.}(2022){Castignani}, {Combes}, {Jablonka},
  {Finn}, {Rudnick}, {Vulcani}, {Desai}, {Zaritsky}, \& {Salom{\'e}}}]{casti22}
{Castignani}, G., {Combes}, F., {Jablonka}, P., {et~al.} 2022, \aap, 657, A9

\bibitem[{{Cautun} {et~al.}(2013{\natexlab{a}}){Cautun}, {van de Weygaert}, \&
  {Jones}}]{Cautun12}
{Cautun}, M., {van de Weygaert}, R., \& {Jones}, B. J.~T. 2013{\natexlab{a}},
  \mnras, 429, 1286

\bibitem[{{Cautun} {et~al.}(2013{\natexlab{b}}){Cautun}, {van de Weygaert}, \&
  {Jones}}]{Cautun13}
{Cautun}, M., {van de Weygaert}, R., \& {Jones}, B. J.~T. 2013{\natexlab{b}},
  \mnras, 429, 1286

\bibitem[{{Cautun} {et~al.}(2014{\natexlab{a}}){Cautun}, {van de Weygaert},
  {Jones}, \& {Frenk}}]{Cautun14}
{Cautun}, M., {van de Weygaert}, R., {Jones}, B. J.~T., \& {Frenk}, C.~S.
  2014{\natexlab{a}}, \mnras, 441, 2923

\bibitem[{{Cautun} {et~al.}(2014{\natexlab{b}}){Cautun}, {van de Weygaert},
  {Jones}, \& {Frenk}}]{Cautun2014}
{Cautun}, M., {van de Weygaert}, R., {Jones}, B. J.~T., \& {Frenk}, C.~S.
  2014{\natexlab{b}}, \mnras, 441, 2923

\bibitem[{{Chen} {et~al.}(2017){Chen}, {Ho}, {Mandelbaum}, {Bahcall},
  {Brownstein}, {Freeman}, {Genovese}, {Schneider}, \& {Wasserman}}]{chen17}
{Chen}, Y.-C., {Ho}, S., {Mandelbaum}, R., {et~al.} 2017, \mnras, 466, 1880

\bibitem[{{Chon} {et~al.}(2014){Chon}, {B{\"o}hringer}, {Collins}, \&
  {Krause}}]{Chon14}
{Chon}, G., {B{\"o}hringer}, H., {Collins}, C.~A., \& {Krause}, M. 2014, \aap,
  567, A144

\bibitem[{{Chung} {et~al.}(2021){Chung}, {Kim}, {Rey}, \& {Lee}}]{Chung21}
{Chung}, J., {Kim}, S., {Rey}, S.-C., \& {Lee}, Y. 2021, \apj, 923, 235

\bibitem[{{Colless} {et~al.}(2001){Colless}, {Dalton}, {Maddox}, {Sutherland},
  {Norberg}, {Cole}, {Bland-Hawthorn}, {Bridges}, {Cannon}, {Collins}, {Couch},
  {Cross}, {Deeley}, {De Propris}, {Driver}, {Efstathiou}, {Ellis}, {Frenk},
  {Glazebrook}, {Jackson}, {Lahav}, {Lewis}, {Lumsden}, {Madgwick}, {Peacock},
  {Peterson}, {Price}, {Seaborne}, \& {Taylor}}]{Colless01}
{Colless}, M., {Dalton}, G., {Maddox}, S., {et~al.} 2001, \mnras, 328, 1039

\bibitem[{{Cornwell} {et~al.}(2024){Cornwell}, {Kuchner}, {Gray},
  {Arag{\'o}n-Salamanca}, {Pearce}, {Cui}, \& {Knebe}}]{Cornwell2024}
{Cornwell}, D.~J., {Kuchner}, U., {Gray}, M.~E., {et~al.} 2024, \mnras, 527, 23

\bibitem[{{Courtois} {et~al.}(2013){Courtois}, {Pomar{\`e}de}, {Tully},
  {Hoffman}, \& {Courtois}}]{Courtois13}
{Courtois}, H.~M., {Pomar{\`e}de}, D., {Tully}, R.~B., {Hoffman}, Y., \&
  {Courtois}, D. 2013, \aj, 146, 69

\bibitem[{{Cybulski} {et~al.}(2014){Cybulski}, {Yun}, {Fazio}, \&
  {Gutermuth}}]{Cybul14}
{Cybulski}, R., {Yun}, M.~S., {Fazio}, G.~G., \& {Gutermuth}, R.~A. 2014,
  \mnras, 439, 3564

\bibitem[{{D{\'a}lya} {et~al.}(2018){D{\'a}lya}, {Galg{\'o}czi}, {Dobos},
  {Frei}, {Heng}, {Macas}, {Messenger}, {Raffai}, \& {de Souza}}]{dalya18}
{D{\'a}lya}, G., {Galg{\'o}czi}, G., {Dobos}, L., {et~al.} 2018, \mnras, 479,
  2374

\bibitem[{{de Vaucouleurs}(1953)}]{devauc53}
{de Vaucouleurs}, G. 1953, \aj, 58, 30

\bibitem[{{de Vaucouleurs}(1956)}]{deVauc56}
{de Vaucouleurs}, G. 1956, Vistas in Astronomy, 2, 1584

\bibitem[{{de Vaucouleurs}(1958)}]{deVauc58}
{de Vaucouleurs}, G. 1958, \nat, 182, 1478

\bibitem[{{de Vaucouleurs}(1975)}]{devauc75}
{de Vaucouleurs}, G. 1975, in Galaxies and the Universe, ed. A.~{Sandage},
  M.~{Sandage}, \& J.~{Kristian} (Chicago Press), 557

\bibitem[{{DESI Collaboration} {et~al.}(2023){DESI Collaboration}, {Adame},
  {Aguilar}, {Ahlen}, {Alam}, {Aldering}, {Alexander}, {Alfarsy}, {Allende
  Prieto}, {Alvarez}, {Alves}, {Anand}, {Andrade-Oliveira}, {Armengaud},
  {Asorey}, {Avila}, {Aviles}, {Bailey}, {Balaguera-Antol{\'\i}nez},
  {Ballester}, {Baltay}, {Bault}, {Bautista}, {Behera}, {Beltran}, {BenZvi},
  {Beraldo e Silva}, {Bermejo-Climent}, {Berti}, {Besuner}, {Beutler},
  {Bianchi}, {Blake}, {Blum}, {Bolton}, {Brieden}, {Brodzeller}, {Brooks},
  {Brown}, {Buckley-Geer}, {Burtin}, {Cabayol-Garcia}, {Cai}, {Canning},
  {Cardiel-Sas}, {Carnero Rosell}, {Castander}, {Cervantes-Cota}, {Chabanier},
  {Chaussidon}, {Chaves-Montero}, {Chen}, {Chuang}, {Claybaugh}, {Cole},
  {Cooper}, {Cuceu}, {Davis}, {Dawson}, {de Belsunce}, {de la Cruz}, {de la
  Macorra}, {de Mattia}, {Demina}, {Demirbozan}, {DeRose}, {Dey}, {Dey},
  {Dhungana}, {Ding}, {Ding}, {Doel}, {Doshi}, {Douglass}, {Edge},
  {Eftekharzadeh}, {Eisenstein}, {Elliott}, {Escoffier}, {Fagrelius}, {Fan},
  {Fanning}, {Fawcett}, {Ferraro}, {Ereza}, {Flaugher}, {Font-Ribera},
  {Forero-S{\'a}nchez}, {Forero-Romero}, {Frenk}, {G{\"a}nsicke},
  {Garc{\'\i}a}, {Garc{\'\i}a-Bellido}, {Garcia-Quintero}, {Garrison},
  {Gil-Mar{\'\i}n}, {Golden-Marx}, {Gontcho}, {Gonzalez-Morales},
  {Gonzalez-Perez}, {Gordon}, {Graur}, {Green}, {Gruen}, {Guy}, {Hadzhiyska},
  {Hahn}, {Han}, {Hanif}, {Herrera-Alcantar}, {Honscheid}, {Hou}, {Howlett},
  {Huterer}, {Ir{\v{s}}i{\v{c}}}, {Ishak}, {Jacques}, {Jana}, {Jiang},
  {Jimenez}, {Jing}, {Joudaki}, {Jullo}, {Juneau}, {Kizhuprakkat},
  {Kara{\c{c}}ayl{\i}}, {Karim}, {Kehoe}, {Kent}, {Khederlarian}, {Kim},
  {Kirkby}, {Kisner}, {Kitaura}, {Kneib}, {Koposov}, {Kov{\'a}cs}, {Kremin},
  {Krolewski}, {L'Huillier}, {Lambert}, {Lamman}, {Lan}, {Landriau}, {Lang},
  {Lange}, {Lasker}, {Le Guillou}, {Leauthaud}, {Levi}, {Li}, {Linder},
  {Lyons}, {Magneville}, {Manera}, {Manser}, {Margala}, {Martini}, {McDonald},
  {Medina}, {Medina-Varela}, {Meisner}, {Mena-Fern{\'a}ndez}, {Meneses-Rizo},
  {Mezcua}, {Miquel}, {Montero-Camacho}, {Moon}, {Moore}, {Moustakas},
  {Mueller}, {Mundet}, {Mu{\~n}oz-Guti{\'e}rrez}, {Myers}, {Nadathur},
  {Napolitano}, {Neveux}, {Newman}, {Nie}, {Nikutta}, {Niz}, {Norberg},
  {Noriega}, {Paillas}, {Palanque-Delabrouille}, {Palmese}, {Zhiwei},
  {Parkinson}, {Penmetsa}, {Percival}, {P{\'e}rez-Fern{\'a}ndez},
  {P{\'e}rez-R{\`a}fols}, {Pieri}, {Poppett}, {Porredon}, {Pothier}, {Prada},
  {Pucha}, {Raichoor}, {Ram{\'\i}rez-P{\'e}rez}, {Ramirez-Solano},
  {Rashkovetskyi}, {Ravoux}, {Rocher}, {Rockosi}, {Ross}, {Rossi}, {Ruggeri},
  {Ruhlmann-Kleider}, {Sabiu}, {Said}, {Saintonge}, {Samushia}, {Sanchez},
  {Saulder}, {Schaan}, {Schlafly}, {Schlegel}, {Scholte}, {Schubnell}, {Seo},
  {Shafieloo}, {Sharples}, {Sheu}, {Silber}, {Sinigaglia}, {Siudek}, {Slepian},
  {Smith}, {Sprayberry}, {Stephey}, {Su{\'a}rez-P{\'e}rez}, {Sun}, {Tan},
  {Tarl{\'e}}, {Tojeiro}, {Ure{\~n}a-L{\'o}pez}, {Vaisakh}, {Valcin}, {Valdes},
  {Valluri}, {Vargas-Maga{\~n}a}, {Variu}, {Verde}, {Walther}, {Wang}, {Wang},
  {Weaver}, {Weaverdyck}, {Wechsler}, {White}, {Xie}, {Yang}, {Y{\`e}che},
  {Yu}, {Yuan}, {Zhang}, {Zhang}, {Zhao}, {Zheng}, {Zhou}, {Zhou}, {Zou},
  {Zou}, \& {Zu}}]{Desi}
{DESI Collaboration}, {Adame}, A.~G., {Aguilar}, J., {et~al.} 2023, arXiv
  e-prints, arXiv:2306.06308

\bibitem[{{Dressler}(1980)}]{dressler1980}
{Dressler}, A. 1980, \apj, 236, 351

\bibitem[{{Drinkwater} {et~al.}(2000){Drinkwater}, {Phillipps}, {Jones},
  {Gregg}, {Deady}, {Davies}, {Parker}, {Sadler}, \& {Smith}}]{Drinkwater2000}
{Drinkwater}, M.~J., {Phillipps}, S., {Jones}, J.~B., {et~al.} 2000, \aap, 355,
  900

\bibitem[{{Ducoin} {et~al.}(2020){Ducoin}, {Corre}, {Leroy}, \& {Le
  Floch}}]{Ducoin20}
{Ducoin}, J.~G., {Corre}, D., {Leroy}, N., \& {Le Floch}, E. 2020, \mnras, 492,
  4768

\bibitem[{{Einasto} {et~al.}(1983){Einasto}, {Corwin}, {Huchra}, {Miller}, \&
  {Tarenghi}}]{einasto1983}
{Einasto}, J., {Corwin}, H.~G., J., {Huchra}, J., {Miller}, R.~H., \&
  {Tarenghi}, M. 1983, Highlights of Astronomy, 6, 757

\bibitem[{{Einasto} {et~al.}(2003){Einasto}, {H{\"u}tsi}, {Einasto}, {Saar},
  {Tucker}, {M{\"u}ller}, {Hein{\"a}m{\"a}ki}, \& {Allam}}]{einasto03}
{Einasto}, J., {H{\"u}tsi}, G., {Einasto}, M., {et~al.} 2003, \aap, 405, 425

\bibitem[{{Einasto} {et~al.}(1980){Einasto}, {Joeveer}, \& {Saar}}]{Einasto08}
{Einasto}, J., {Joeveer}, M., \& {Saar}, E. 1980, \mnras, 193, 353

\bibitem[{{Einasto}(1992)}]{einasto92}
{Einasto}, M. 1992, \mnras, 258, 571

\bibitem[{{Einasto} \& {Einasto}(1987)}]{einasto87}
{Einasto}, M. \& {Einasto}, J. 1987, \mnras, 226, 543

\bibitem[{{Einasto} {et~al.}(2024){Einasto}, {Einasto}, {Tenjes}, {Korhonen},
  {Kipper}, {Tempel}, {Liivam{\"a}gi}, \& {Hein{\"a}m{\"a}ki}}]{einasto24}
{Einasto}, M., {Einasto}, J., {Tenjes}, P., {et~al.} 2024, \aap, 681, A91

\bibitem[{{Einasto} {et~al.}(2018){Einasto}, {Gramann}, {Park}, {Kim},
  {Deshev}, {Tempel}, {Hein{\"a}m{\"a}ki}, {Lietzen}, {L{\"a}hteenm{\"a}ki},
  {Einasto}, \& {Saar}}]{Einasto18}
{Einasto}, M., {Gramann}, M., {Park}, C., {et~al.} 2018, \aap, 620, A149

\bibitem[{{Einasto} {et~al.}(2021){Einasto}, {Kipper}, {Tenjes}, {Lietzen},
  {Tempel}, {Liivam{\"a}gi}, {Einasto}, {Tamm}, {Hein{\"a}m{\"a}ki}, \&
  {Nurmi}}]{Einasto21}
{Einasto}, M., {Kipper}, R., {Tenjes}, P., {et~al.} 2021, \aap, 649, A51

\bibitem[{{Einasto} {et~al.}(2017){Einasto}, {Lietzen}, {Gramann}, {Saar},
  {Tempel}, {Liivam{\"a}gi}, {Montero-Dorta}, {Streblyanska}, {Maraston}, \&
  {Rubi{\~n}o-Mart{\'\i}n}}]{Einasto17}
{Einasto}, M., {Lietzen}, H., {Gramann}, M., {et~al.} 2017, \aap, 603, A5

\bibitem[{{Einasto} {et~al.}(2016){Einasto}, {Lietzen}, {Gramann}, {Tempel},
  {Saar}, {Liivam{\"a}gi}, {Hein{\"a}m{\"a}ki}, {Nurmi}, \&
  {Einasto}}]{einasto16}
{Einasto}, M., {Lietzen}, H., {Gramann}, M., {et~al.} 2016, \aap, 595, A70

\bibitem[{{Einasto} {et~al.}(2011{\natexlab{a}}){Einasto}, {Liivam{\"a}gi},
  {Tago}, {Saar}, {Tempel}, {Einasto}, {Mart{\'\i}nez}, \&
  {Hein{\"a}m{\"a}ki}}]{einasto11}
{Einasto}, M., {Liivam{\"a}gi}, L.~J., {Tago}, E., {et~al.} 2011{\natexlab{a}},
  \aap, 532, A5

\bibitem[{{Einasto} {et~al.}(2011{\natexlab{b}}){Einasto}, {Liivam{\"a}gi},
  {Tempel}, {Saar}, {Tago}, {Einasto}, {Enkvist}, {Einasto}, {Mart{\'\i}nez},
  {Hein{\"a}m{\"a}ki}, \& {Nurmi}}]{meinasto11}
{Einasto}, M., {Liivam{\"a}gi}, L.~J., {Tempel}, E., {et~al.}
  2011{\natexlab{b}}, \apj, 736, 51

\bibitem[{{Einasto} {et~al.}(1997){Einasto}, {Tago}, {Jaaniste}, {Einasto}, \&
  {Andernach}}]{einasto97}
{Einasto}, M., {Tago}, E., {Jaaniste}, J., {Einasto}, J., \& {Andernach}, H.
  1997, \aaps, 123, 119

\bibitem[{{Einasto} {et~al.}(2010){Einasto}, {Tago}, {Saar}, {Nurmi},
  {Enkvist}, {Einasto}, {Hein{\"a}m{\"a}ki}, {Liivam{\"a}gi}, {Tempel},
  {Einasto}, {Mart{\'\i}nez}, {Vennik}, \& {Pihajoki}}]{einasto10}
{Einasto}, M., {Tago}, E., {Saar}, E., {et~al.} 2010, \aap, 522, A92

\bibitem[{{Einasto} {et~al.}(2022){Einasto}, {Tenjes}, {Gramann}, {Lietzen},
  {Kipper}, {Liivam{\"a}gi}, {Tempel}, {Sankhyayan}, \& {Einasto}}]{einasto22}
{Einasto}, M., {Tenjes}, P., {Gramann}, M., {et~al.} 2022, \aap, 666, A52

\bibitem[{{Euclid Collaboration} {et~al.}(2022){Euclid Collaboration},
  {Scaramella}, {Amiaux}, {Mellier}, {Burigana}, {Carvalho}, {Cuillandre}, {Da
  Silva}, {Derosa}, {Dinis}, {Maiorano}, {Maris}, {Tereno}, {Laureijs},
  {Boenke}, {Buenadicha}, {Dupac}, {Gaspar Venancio}, {G{\'o}mez-{\'A}lvarez},
  {Hoar}, {Lorenzo Alvarez}, {Racca}, {Saavedra-Criado}, {Schwartz}, {Vavrek},
  {Schirmer}, {Aussel}, {Azzollini}, {Cardone}, {Cropper}, {Ealet}, {Garilli},
  {Gillard}, {Granett}, {Guzzo}, {Hoekstra}, {Jahnke}, {Kitching}, {Maciaszek},
  {Meneghetti}, {Miller}, {Nakajima}, {Niemi}, {Pasian}, {Percival},
  {Pottinger}, {Sauvage}, {Scodeggio}, {Wachter}, {Zacchei}, {Aghanim},
  {Amara}, {Auphan}, {Auricchio}, {Awan}, {Balestra}, {Bender}, {Bodendorf},
  {Bonino}, {Branchini}, {Brau-Nogue}, {Brescia}, {Candini}, {Capobianco},
  {Carbone}, {Carlberg}, {Carretero}, {Casas}, {Castander}, {Castellano},
  {Cavuoti}, {Cimatti}, {Cledassou}, {Congedo}, {Conselice}, {Conversi},
  {Copin}, {Corcione}, {Costille}, {Courbin}, {Degaudenzi}, {Douspis},
  {Dubath}, {Duncan}, {Dusini}, {Farrens}, {Ferriol}, {Fosalba}, {Fourmanoit},
  {Frailis}, {Franceschi}, {Franzetti}, {Fumana}, {Gillis}, {Giocoli},
  {Grazian}, {Grupp}, {Haugan}, {Holmes}, {Hormuth}, {Hudelot}, {Kermiche},
  {Kiessling}, {Kilbinger}, {Kohley}, {Kubik}, {K{\"u}mmel}, {Kunz},
  {Kurki-Suonio}, {Lahav}, {Ligori}, {Lilje}, {Lloro}, {Mansutti}, {Marggraf},
  {Markovic}, {Marulli}, {Massey}, {Maurogordato}, {Melchior}, {Merlin},
  {Meylan}, {Mohr}, {Moresco}, {Morin}, {Moscardini}, {Munari}, {Nichol},
  {Padilla}, {Paltani}, {Peacock}, {Pedersen}, {Pettorino}, {Pires}, {Poncet},
  {Popa}, {Pozzetti}, {Raison}, {Rebolo}, {Rhodes}, {Rix}, {Roncarelli},
  {Rossetti}, {Saglia}, {Schneider}, {Schrabback}, {Secroun}, {Seidel},
  {Serrano}, {Sirignano}, {Sirri}, {Skottfelt}, {Stanco}, {Starck},
  {Tallada-Cresp{\'\i}}, {Tavagnacco}, {Taylor}, {Teplitz}, {Toledo-Moreo},
  {Torradeflot}, {Trifoglio}, {Valentijn}, {Valenziano}, {Verdoes Kleijn},
  {Wang}, {Welikala}, {Weller}, {Wetzstein}, {Zamorani}, {Zoubian}, {Andreon},
  {Baldi}, {Bardelli}, {Boucaud}, {Camera}, {Di Ferdinando}, {Fabbian},
  {Farinelli}, {Galeotta}, {Graci{\'a}-Carpio}, {Maino}, {Medinaceli}, {Mei},
  {Neissner}, {Polenta}, {Renzi}, {Romelli}, {Rosset}, {Sureau}, {Tenti},
  {Vassallo}, {Zucca}, {Baccigalupi}, {Balaguera-Antol{\'\i}nez}, {Battaglia},
  {Biviano}, {Borgani}, {Bozzo}, {Cabanac}, {Cappi}, {Casas}, {Castignani},
  {Colodro-Conde}, {Coupon}, {Courtois}, {Cuby}, {de la Torre}, {Desai},
  {Dole}, {Fabricius}, {Farina}, {Ferreira}, {Finelli}, {Flose-Reimberg},
  {Fotopoulou}, {Ganga}, {Gozaliasl}, {Hook}, {Keihanen}, {Kirkpatrick},
  {Liebing}, {Lindholm}, {Mainetti}, {Martinelli}, {Martinet}, {Maturi},
  {McCracken}, {Metcalf}, {Morgante}, {Nightingale}, {Nucita}, {Patrizii},
  {Potter}, {Riccio}, {S{\'a}nchez}, {Sapone}, {Schewtschenko}, {Schultheis},
  {Scottez}, {Teyssier}, {Tutusaus}, {Valiviita}, {Viel}, {Vriend}, \&
  {Whittaker}}]{euclid2022}
{Euclid Collaboration}, {Scaramella}, R., {Amiaux}, J., {et~al.} 2022, \aap,
  662, A112

\bibitem[{{Fairall} {et~al.}(1994){Fairall}, {Paverd}, \& {Ashley}}]{Fairall94}
{Fairall}, A.~P., {Paverd}, W.~R., \& {Ashley}, R.~P. 1994, in Astronomical
  Society of the Pacific Conference Series, Vol.~67, Unveiling Large-Scale
  Structures Behind the Milky Way, ed. C.~{Balkowski} \& R.~C.
  {Kraan-Korteweg}, 21

\bibitem[{{For} {et~al.}(2023){For}, {Spekkens}, {Staveley-Smith}, {Bekki},
  {Karunakaran}, {Catinella}, {Koribalski}, {Lee-Waddell}, {Madrid},
  {Murugeshan}, {Rhee}, {Westmeier}, {Wong}, {Zaritsky}, \&
  {Donnerstein}}]{For23}
{For}, B.~Q., {Spekkens}, K., {Staveley-Smith}, L., {et~al.} 2023, \mnras, 526,
  3130

\bibitem[{{Fujita}(2004)}]{fujita04}
{Fujita}, Y. 2004, \pasj, 56, 29

\bibitem[{{Gal{\'a}rraga-Espinosa} {et~al.}(2023){Gal{\'a}rraga-Espinosa},
  {Garaldi}, \& {Kauffmann}}]{espinosa23}
{Gal{\'a}rraga-Espinosa}, D., {Garaldi}, E., \& {Kauffmann}, G. 2023, \aap,
  671, A160

\bibitem[{{Geller} \& {Huchra}(1989)}]{Geller89}
{Geller}, M.~J. \& {Huchra}, J.~P. 1989, Science, 246, 897

\bibitem[{{Gott} {et~al.}(2005){Gott}, {Juri{\'c}}, {Schlegel}, {Hoyle},
  {Vogeley}, {Tegmark}, {Bahcall}, \& {Brinkmann}}]{Gott05}
{Gott}, J.~Richard, I., {Juri{\'c}}, M., {Schlegel}, D., {et~al.} 2005, \apj,
  624, 463

\bibitem[{{Gunn} \& {Gott}(1972)}]{gunn72}
{Gunn}, J.~E. \& {Gott}, J.~Richard, I. 1972, \apj, 176, 1

\bibitem[{{Haines} {et~al.}(2018){Haines}, {Busarello}, {Merluzzi}, {Pimbblet},
  {Vogt}, {Dopita}, {Mercurio}, {Grado}, \& {Limatola}}]{Haines18}
{Haines}, C.~P., {Busarello}, G., {Merluzzi}, P., {et~al.} 2018, \mnras, 481,
  1055

\bibitem[{{Haines} {et~al.}(2006){Haines}, {Merluzzi}, {Mercurio}, {Gargiulo},
  {Krusanova}, {Busarello}, {La Barbera}, \& {Capaccioli}}]{haines06}
{Haines}, C.~P., {Merluzzi}, P., {Mercurio}, A., {et~al.} 2006, \mnras, 371, 55

\bibitem[{{Hatamkhani} {et~al.}(2023){Hatamkhani}, {Kraan-Korteweg}, {Blyth},
  {Said}, \& {Elagali}}]{hatam23}
{Hatamkhani}, N., {Kraan-Korteweg}, R.~C., {Blyth}, S.~L., {Said}, K., \&
  {Elagali}, A. 2023, \mnras, 522, 2223

\bibitem[{{He{\ss}} {et~al.}(2013){He{\ss}}, {Kitaura}, \&
  {Gottl{\"o}ber}}]{Hes13}
{He{\ss}}, S., {Kitaura}, F.-S., \& {Gottl{\"o}ber}, S. 2013, \mnras, 435, 2065

\bibitem[{{Hoosain} {et~al.}(2024){Hoosain}, {Blyth}, {Skelton}, {Kannappan},
  {Stark}, {Eckert}, {Hutchens}, {Carr}, \& {Kraljic}}]{Hoosain24}
{Hoosain}, M., {Blyth}, S.-L., {Skelton}, R.~E., {et~al.} 2024, \mnras, 528,
  4139

\bibitem[{{Huchra} \& {Geller}(1982)}]{huchra82}
{Huchra}, J.~P. \& {Geller}, M.~J. 1982, \apj, 257, 423

\bibitem[{{Huchra} {et~al.}(2012){Huchra}, {Macri}, {Masters}, {Jarrett},
  {Berlind}, {Calkins}, {Crook}, {Cutri}, {Erdo{\v{g}}du}, {Falco}, {George},
  {Hutcheson}, {Lahav}, {Mader}, {Mink}, {Martimbeau}, {Schneider},
  {Skrutskie}, {Tokarz}, \& {Westover}}]{huchra12}
{Huchra}, J.~P., {Macri}, L.~M., {Masters}, K.~L., {et~al.} 2012, \apjs, 199,
  26

\bibitem[{Hunter(2007)}]{Hunter:2007}
Hunter, J.~D. 2007, Computing in Science \& Engineering, 9, 90

\bibitem[{{Icke}(1973)}]{Icke1973}
{Icke}, V. 1973, \aap, 27, 1

\bibitem[{Inc.(2015)}]{plotly}
Inc., P.~T. 2015, Collaborative data science

\bibitem[{{J{\~o}eveer} {et~al.}(1978){J{\~o}eveer}, {Einasto}, \&
  {Tago}}]{joeveer78}
{J{\~o}eveer}, M., {Einasto}, J., \& {Tago}, E. 1978, \mnras, 185, 357

\bibitem[{{Jones} {et~al.}(2009){Jones}, {Read}, {Saunders}, {Colless},
  {Jarrett}, {Parker}, {Fairall}, {Mauch}, {Sadler}, {Watson}, {Burton},
  {Campbell}, {Cass}, {Croom}, {Dawe}, {Fiegert}, {Frankcombe}, {Hartley},
  {Huchra}, {James}, {Kirby}, {Lahav}, {Lucey}, {Mamon}, {Moore}, {Peterson},
  {Prior}, {Proust}, {Russell}, {Safouris}, {Wakamatsu}, {Westra}, \&
  {Williams}}]{Jones09}
{Jones}, D.~H., {Read}, M.~A., {Saunders}, W., {et~al.} 2009, \mnras, 399, 683

\bibitem[{{Jones} \& {Jones}(1980)}]{JonesJones1980}
{Jones}, J.~E. \& {Jones}, B.~J.~T. 1980, \mnras, 191, 685

\bibitem[{{Kauffmann} {et~al.}(2004){Kauffmann}, {White}, {Heckman},
  {M{\'e}nard}, {Brinchmann}, {Charlot}, {Tremonti}, \&
  {Brinkmann}}]{kauffmann04}
{Kauffmann}, G., {White}, S. D.~M., {Heckman}, T.~M., {et~al.} 2004, \mnras,
  353, 713

\bibitem[{{Kitaura}(2013)}]{Kitaura13}
{Kitaura}, F.~S. 2013, \mnras, 429, L84

\bibitem[{{Kitaura} {et~al.}(2024){Kitaura}, {Sinigaglia},
  {Balaguera-Antol{\'\i}nez}, \& {Favole}}]{Kitaura23}
{Kitaura}, F.~S., {Sinigaglia}, F., {Balaguera-Antol{\'\i}nez}, A., \&
  {Favole}, G. 2024, \aap, 683, A215

\bibitem[{{Kleiner} {et~al.}(2021){Kleiner}, {Serra}, {Maccagni}, {Venhola},
  {Morokuma-Matsui}, {Peletier}, {Iodice}, {Raj}, {de Blok}, {Comrie},
  {J{\'o}zsa}, {Kamphuis}, {Loni}, {Loubser}, {Moln{\'a}r}, {Passmoor},
  {Ramatsoku}, {Sivitilli}, {Smirnov}, {Thorat}, \& {Vitello}}]{kleiner21}
{Kleiner}, D., {Serra}, P., {Maccagni}, F.~M., {et~al.} 2021, \aap, 648, A32

\bibitem[{{Kourkchi} \& {Tully}(2017)}]{Kourkchi2017}
{Kourkchi}, E. \& {Tully}, R.~B. 2017, \apj, 843, 16

\bibitem[{{Kraan-Korteweg} {et~al.}(2017){Kraan-Korteweg}, {Cluver}, {Bilicki},
  {Jarrett}, {Colless}, {Elagali}, {B{\"o}hringer}, \& {Chon}}]{kraan17}
{Kraan-Korteweg}, R.~C., {Cluver}, M.~E., {Bilicki}, M., {et~al.} 2017, \mnras,
  466, L29

\bibitem[{{Kraljic} {et~al.}(2018){Kraljic}, {Arnouts}, {Pichon}, {Laigle}, {de
  la Torre}, {Vibert}, {Cadiou}, {Dubois}, {Treyer}, {Schimd}, {Codis}, {de
  Lapparent}, {Devriendt}, {Hwang}, {Le Borgne}, {Malavasi}, {Milliard},
  {Musso}, {Pogosyan}, {Alpaslan}, {Bland-Hawthorn}, \& {Wright}}]{kraljic18}
{Kraljic}, K., {Arnouts}, S., {Pichon}, C., {et~al.} 2018, \mnras, 474, 547

\bibitem[{{Kuchner} {et~al.}(2022){Kuchner}, {Haggar}, {Arag{\'o}n-Salamanca},
  {Pearce}, {Gray}, {Rost}, {Cui}, {Knebe}, \& {Yepes}}]{Kuchner2022}
{Kuchner}, U., {Haggar}, R., {Arag{\'o}n-Salamanca}, A., {et~al.} 2022, \mnras,
  510, 581

\bibitem[{{Kuutma} {et~al.}(2017){Kuutma}, {Tamm}, \& {Tempel}}]{kuutuma17}
{Kuutma}, T., {Tamm}, A., \& {Tempel}, E. 2017, \aap, 600, L6

\bibitem[{{Lavaux} \& {Hudson}(2011)}]{Lavaux11}
{Lavaux}, G. \& {Hudson}, M.~J. 2011, \mnras, 416, 2840

\bibitem[{{Libeskind} {et~al.}(2014){Libeskind}, {Hoffman}, \&
  {Gottl{\"o}ber}}]{libeskind14}
{Libeskind}, N.~I., {Hoffman}, Y., \& {Gottl{\"o}ber}, S. 2014, \mnras, 441,
  1974

\bibitem[{{Libeskind} {et~al.}(2018){Libeskind}, {van de Weygaert}, {Cautun},
  {Falck}, {Tempel}, {Abel}, {Alpaslan}, {Arag{\'o}n-Calvo}, {Forero-Romero},
  {Gonzalez}, {Gottl{\"o}ber}, {Hahn}, {Hellwing}, {Hoffman}, {Jones},
  {Kitaura}, {Knebe}, {Manti}, {Neyrinck}, {Nuza}, {Padilla}, {Platen},
  {Ramachandra}, {Robotham}, {Saar}, {Shandarin}, {Steinmetz}, {Stoica},
  {Sousbie}, \& {Yepes}}]{Libeskind2018}
{Libeskind}, N.~I., {van de Weygaert}, R., {Cautun}, M., {et~al.} 2018, \mnras,
  473, 1195

\bibitem[{{Lietzen} {et~al.}(2016){Lietzen}, {Tempel}, {Liivam{\"a}gi},
  {Montero-Dorta}, {Einasto}, {Streblyanska}, {Maraston},
  {Rubi{\~n}o-Mart{\'\i}n}, \& {Saar}}]{lietzen16}
{Lietzen}, H., {Tempel}, E., {Liivam{\"a}gi}, L.~J., {et~al.} 2016, \aap, 588,
  L4

\bibitem[{{Liivam{\"a}gi} {et~al.}(2012){Liivam{\"a}gi}, {Tempel}, \&
  {Saar}}]{liivam12}
{Liivam{\"a}gi}, L.~J., {Tempel}, E., \& {Saar}, E. 2012, \aap, 539, A80

\bibitem[{{Lopes} {et~al.}(2024){Lopes}, {Ribeiro}, \& {Brambila}}]{Lopes2024}
{Lopes}, P. A.~A., {Ribeiro}, A. L.~B., \& {Brambila}, D. 2024, \mnras, 527,
  L19

\bibitem[{{Makarov} \& {Karachentsev}(2011)}]{Makarov2011}
{Makarov}, D. \& {Karachentsev}, I. 2011, \mnras, 412, 2498

\bibitem[{{Malavasi} {et~al.}(2023){Malavasi}, {Sorce}, {Dolag}, \&
  {Aghanim}}]{Malavasi23}
{Malavasi}, N., {Sorce}, J.~G., {Dolag}, K., \& {Aghanim}, N. 2023, \aap, 675,
  A76

\bibitem[{{Marasco} {et~al.}(2023){Marasco}, {Poggianti}, {Fritz}, {Werle},
  {Vulcani}, {Moretti}, {Gullieuszik}, \& {Kulier}}]{marasco23}
{Marasco}, A., {Poggianti}, B.~M., {Fritz}, J., {et~al.} 2023, \mnras, 525,
  5359

\bibitem[{{Mei} {et~al.}(2023){Mei}, {Hatch}, {Amodeo}, {Afanasiev}, {De
  Breuck}, {Stern}, {Cooke}, {Gonzalez}, {Noirot}, {Rettura}, {Seymour},
  {Stanford}, {Vernet}, \& {Wylezalek}}]{simona23}
{Mei}, S., {Hatch}, N.~A., {Amodeo}, S., {et~al.} 2023, \aap, 670, A58

\bibitem[{Mohammadi \& Bunte(2020)}]{Mohammadi2020_EM3A}
Mohammadi, M. \& Bunte, K. 2020, in Intelligent Data Engineering and Automated
  Learning -- IDEAL 2020, ed. C.~Analide, P.~Novais, D.~Camacho, \& H.~Yin
  (Cham: Springer International Publishing), 12--24

\bibitem[{{Nasonova} {et~al.}(2011){Nasonova}, {de Freitas Pacheco}, \&
  {Karachentsev}}]{nasonova2011}
{Nasonova}, O.~G., {de Freitas Pacheco}, J.~A., \& {Karachentsev}, I.~D. 2011,
  \aap, 532, A104

\bibitem[{{Peebles}(2023)}]{peebles23}
{Peebles}, P.~J.~E. 2023, \mnras, 526, 4490

\bibitem[{{Pfeifer} {et~al.}(2022){Pfeifer}, {Libeskind}, {Hoffman},
  {Hellwing}, {Bilicki}, \& {Naidoo}}]{Pfeifer2022}
{Pfeifer}, S., {Libeskind}, N.~I., {Hoffman}, Y., {et~al.} 2022, \mnras, 514,
  470

\bibitem[{{Phillipps} \& {Davies}(1992)}]{PhillippsDavies1992}
{Phillipps}, S. \& {Davies}, J. 1992, in Astrophysics and Space Science
  Library, Vol. 174, Digitised Optical Sky Surveys, ed. H.~T. {MacGillivray} \&
  E.~B. {Thomson}, 295

\bibitem[{{Rams{\o}y} {et~al.}(2021){Rams{\o}y}, {Slyz}, {Devriendt}, {Laigle},
  \& {Dubois}}]{Ramsoy2021}
{Rams{\o}y}, M., {Slyz}, A., {Devriendt}, J., {Laigle}, C., \& {Dubois}, Y.
  2021, \mnras, 502, 351

\bibitem[{{Rom{\'a}n} \& {Trujillo}(2017)}]{Roman2017b}
{Rom{\'a}n}, J. \& {Trujillo}, I. 2017, \mnras, 468, 703

\bibitem[{{Santiago-Bautista} {et~al.}(2020){Santiago-Bautista}, {Caretta},
  {Bravo-Alfaro}, {Pointecouteau}, \& {Andernach}}]{santiago20}
{Santiago-Bautista}, I., {Caretta}, C.~A., {Bravo-Alfaro}, H., {Pointecouteau},
  E., \& {Andernach}, H. 2020, \aap, 637, A31

\bibitem[{{Sarron} {et~al.}(2019){Sarron}, {Adami}, {Durret}, \&
  {Laigle}}]{sarron19}
{Sarron}, F., {Adami}, C., {Durret}, F., \& {Laigle}, C. 2019, \aap, 632, A49

\bibitem[{{Schaye} {et~al.}(2015){Schaye}, {Crain}, {Bower}, {Furlong},
  {Schaller}, {Theuns}, {Dalla Vecchia}, {Frenk}, {McCarthy}, {Helly},
  {Jenkins}, {Rosas-Guevara}, {White}, {Baes}, {Booth}, {Camps}, {Navarro},
  {Qu}, {Rahmati}, {Sawala}, {Thomas}, \& {Trayford}}]{schaye15}
{Schaye}, J., {Crain}, R.~A., {Bower}, R.~G., {et~al.} 2015, \mnras, 446, 521

\bibitem[{{Shapley}(1930)}]{shapley}
{Shapley}, H. 1930, Harvard College Observatory Bulletin, 874, 9

\bibitem[{{Skrutskie} {et~al.}(2006){Skrutskie}, {Cutri}, {Stiening},
  {Weinberg}, {Schneider}, {Carpenter}, {Beichman}, {Capps}, {Chester},
  {Elias}, {Huchra}, {Liebert}, {Lonsdale}, {Monet}, {Price}, {Seitzer},
  {Jarrett}, {Kirkpatrick}, {Gizis}, {Howard}, {Evans}, {Fowler}, {Fullmer},
  {Hurt}, {Light}, {Kopan}, {Marsh}, {McCallon}, {Tam}, {Van Dyk}, \&
  {Wheelock}}]{2mass}
{Skrutskie}, M.~F., {Cutri}, R.~M., {Stiening}, R., {et~al.} 2006, \aj, 131,
  1163

\bibitem[{{Sorce} {et~al.}(2016){Sorce}, {Gottl{\"o}ber}, {Yepes}, {Hoffman},
  {Courtois}, {Steinmetz}, {Tully}, {Pomar{\`e}de}, \& {Carlesi}}]{Sorce16}
{Sorce}, J.~G., {Gottl{\"o}ber}, S., {Yepes}, G., {et~al.} 2016, \mnras, 455,
  2078

\bibitem[{{Sousbie}(2011)}]{sousbie11}
{Sousbie}, T. 2011, \mnras, 414, 350

\bibitem[{{Springel}(2005)}]{springel05}
{Springel}, V. 2005, \mnras, 364, 1105

\bibitem[{Taghribi {et~al.}(2023)Taghribi, Bunte, Smith, Shin, Mastropietro,
  Peletier, \& Tiňo}]{Taghribi2023}
Taghribi, A., Bunte, K., Smith, R., {et~al.} 2023, IEEE Transactions on
  Knowledge and Data Engineering, 35, 6014

\bibitem[{{Tegmark} {et~al.}(2004){Tegmark}, {Blanton}, {Strauss}, {Hoyle},
  {Schlegel}, {Scoccimarro}, {Vogeley}, {Weinberg}, {Zehavi}, {Berlind},
  {Budavari}, {Connolly}, {Eisenstein}, {Finkbeiner}, {Frieman}, {Gunn},
  {Hamilton}, {Hui}, {Jain}, {Johnston}, {Kent}, {Lin}, {Nakajima}, {Nichol},
  {Ostriker}, {Pope}, {Scranton}, {Seljak}, {Sheth}, {Stebbins}, {Szalay},
  {Szapudi}, {Verde}, {Xu}, {Annis}, {Bahcall}, {Brinkmann}, {Burles},
  {Castander}, {Csabai}, {Loveday}, {Doi}, {Fukugita}, {Gott}, {Hennessy},
  {Hogg}, {Ivezi{\'c}}, {Knapp}, {Lamb}, {Lee}, {Lupton}, {McKay}, {Kunszt},
  {Munn}, {O'Connell}, {Peoples}, {Pier}, {Richmond}, {Rockosi}, {Schneider},
  {Stoughton}, {Tucker}, {Vanden Berk}, {Yanny}, {York}, \& {SDSS
  Collaboration}}]{tegmark04}
{Tegmark}, M., {Blanton}, M.~R., {Strauss}, M.~A., {et~al.} 2004, \apj, 606,
  702

\bibitem[{{Tempel} {et~al.}(2016){Tempel}, {Kipper}, {Tamm}, {Gramann},
  {Einasto}, {Sepp}, \& {Tuvikene}}]{Tempel2016a}
{Tempel}, E., {Kipper}, R., {Tamm}, A., {et~al.} 2016, \aap, 588, A14

\bibitem[{{Tempel} {et~al.}(2014){Tempel}, {Stoica}, {Mart{\'\i}nez},
  {Liivam{\"a}gi}, {Castellan}, \& {Saar}}]{Tempel14}
{Tempel}, E., {Stoica}, R.~S., {Mart{\'\i}nez}, V.~J., {et~al.} 2014, \mnras,
  438, 3465

\bibitem[{{Tully}(1987)}]{Tully1987a}
{Tully}, R.~B. 1987, \apj, 321, 280

\bibitem[{{Tully} {et~al.}(2014){Tully}, {Courtois}, {Hoffman}, \&
  {Pomar{\`e}de}}]{tully14}
{Tully}, R.~B., {Courtois}, H., {Hoffman}, Y., \& {Pomar{\`e}de}, D. 2014,
  \nat, 513, 71

\bibitem[{{Tully} {et~al.}(2013){Tully}, {Courtois}, {Dolphin}, {Fisher},
  {H{\'e}raudeau}, {Jacobs}, {Karachentsev}, {Makarov}, {Makarova},
  {Mitronova}, {Rizzi}, {Shaya}, {Sorce}, \& {Wu}}]{Tully13}
{Tully}, R.~B., {Courtois}, H.~M., {Dolphin}, A.~E., {et~al.} 2013, \aj, 146,
  86

\bibitem[{{Tully} \& {Fisher}(1987)}]{tully87}
{Tully}, R.~B. \& {Fisher}, J.~R. 1987, {Atlas of Nearby Galaxies} (Cambridge
  University Press)

\bibitem[{{Tully} {et~al.}(2009){Tully}, {Rizzi}, {Shaya}, {Courtois},
  {Makarov}, \& {Jacobs}}]{Tully09}
{Tully}, R.~B., {Rizzi}, L., {Shaya}, E.~J., {et~al.} 2009, \aj, 138, 323

\bibitem[{{van de Weygaert} \& {Bond}(2008)}]{vdW08}
{van de Weygaert}, R. \& {Bond}, J.~R. 2008, in A Pan-Chromatic View of
  Clusters of Galaxies and the Large-Scale Structure, ed. M.~{Plionis},
  O.~{L{\'o}pez-Cruz}, \& D.~{Hughes}, Vol. 740 (Springer), 335

\bibitem[{{van de Weygaert} \& {Schaap}(2009)}]{vDW09}
{van de Weygaert}, R. \& {Schaap}, W. 2009, in Data Analysis in Cosmology, ed.
  V.~J. {Mart{\'\i}nez}, E.~{Saar}, E.~{Mart{\'\i}nez-Gonz{\'a}lez}, \& M.~J.
  {Pons-Border{\'\i}a}, Vol. 665 (Springer), 291--413

\bibitem[{{Venturi} {et~al.}(2022){Venturi}, {Giacintucci}, {Merluzzi},
  {Bardelli}, {Busarello}, {Dallacasa}, {Sikhosana}, {Marvil}, {Smirnov},
  {Bourdin}, {Mazzotta}, {Rossetti}, {Rudnick}, {Bernardi}, {Br{\"u}ggen},
  {Carretti}, {Cassano}, {Di Gennaro}, {Gastaldello}, {Kale}, {Knowles},
  {Koribalski}, {Heywood}, {Hopkins}, {Norris}, {Reiprich}, {Tasse},
  {Vernstrom}, {Zucca}, {Bester}, {Diego}, \& {Kanapathippillai}}]{Venturi22}
{Venturi}, T., {Giacintucci}, S., {Merluzzi}, P., {et~al.} 2022, \aap, 660, A81

\bibitem[{{Vogelsberger} {et~al.}(2014){Vogelsberger}, {Genel}, {Springel},
  {Torrey}, {Sijacki}, {Xu}, {Snyder}, {Nelson}, \& {Hernquist}}]{Vogel14}
{Vogelsberger}, M., {Genel}, S., {Springel}, V., {et~al.} 2014, \mnras, 444,
  1518

\bibitem[{{Vulcani} {et~al.}(2015){Vulcani}, {Poggianti}, {Fritz}, {Fasano},
  {Moretti}, {Calvi}, \& {Paccagnella}}]{benedetta15}
{Vulcani}, B., {Poggianti}, B.~M., {Fritz}, J., {et~al.} 2015, \apj, 798, 52

\bibitem[{{White} \& {Rees}(1978)}]{White78}
{White}, S.~D.~M. \& {Rees}, M.~J. 1978, \mnras, 183, 341

\bibitem[{{White} \& {Silk}(1979)}]{WhiteSilk1979}
{White}, S.~D.~M. \& {Silk}, J. 1979, \apj, 231, 1

\bibitem[{{Winkel} {et~al.}(2021){Winkel}, {Pasquali}, {Kraljic}, {Smith},
  {Gallazzi}, \& {Jackson}}]{winkel21}
{Winkel}, N., {Pasquali}, A., {Kraljic}, K., {et~al.} 2021, \mnras, 505, 4920

\bibitem[{{Zabludoff} \& {Mulchaey}(1998)}]{zab98}
{Zabludoff}, A.~I. \& {Mulchaey}, J.~S. 1998, \apj, 496, 39

\bibitem[{{Zeldovich} {et~al.}(1982){Zeldovich}, {Einasto}, \&
  {Shandarin}}]{zeldovich82}
{Zeldovich}, I.~B., {Einasto}, J., \& {Shandarin}, S.~F. 1982, \nat, 300, 407

\bibitem[{{Zel'dovich}(1970)}]{Zeldovich1970}
{Zel'dovich}, Y.~B. 1970, \aap, 5, 84

\end{thebibliography}
\begin{appendix} 
\section{}\label{appendix}
\begin{figure*}
  \centering
   \includegraphics[scale=0.4]{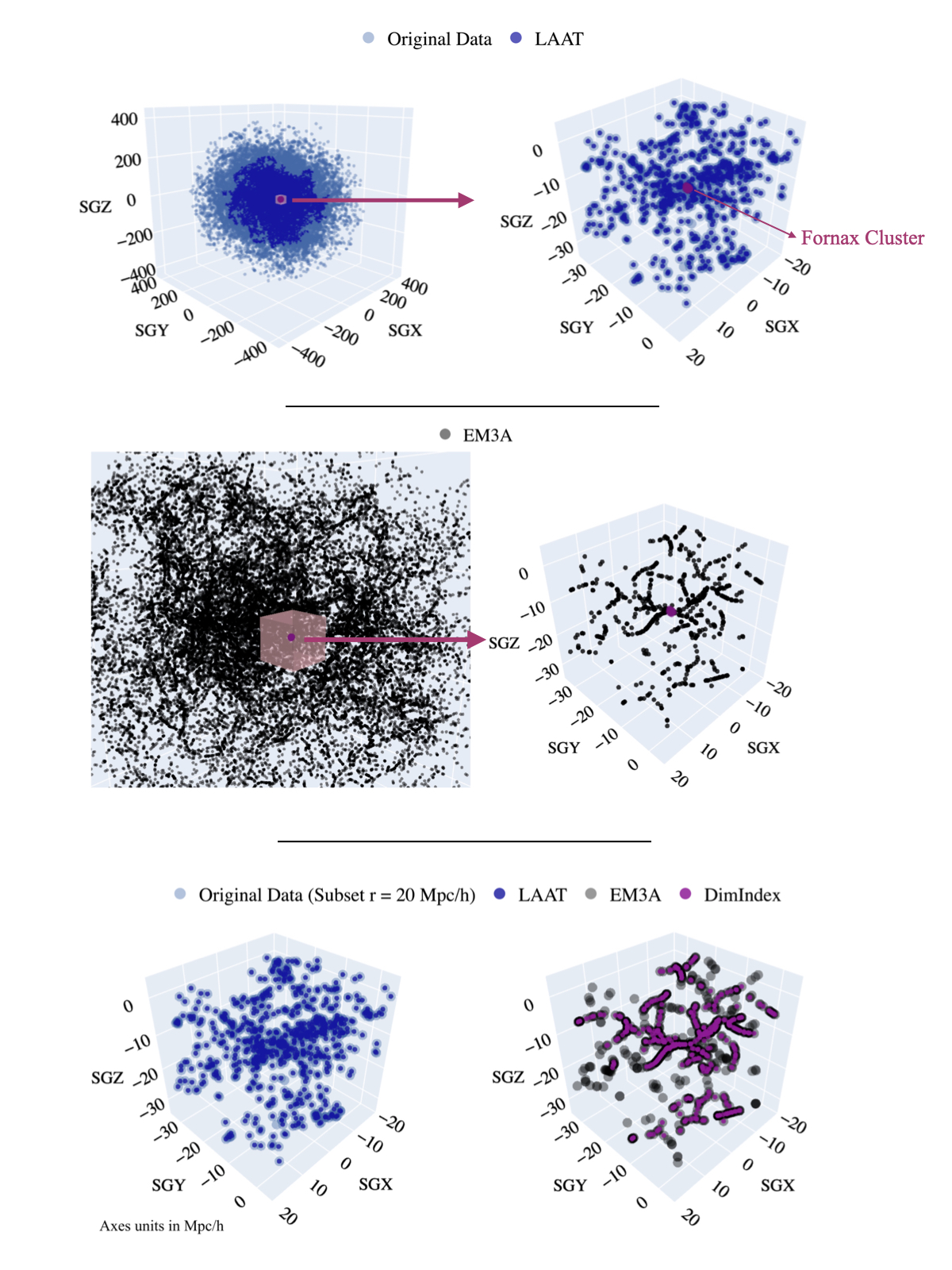}
     
      \caption{Results of LAAT, EM3A, and DimIndex on our data. \textit{Top panel}: 3D representation of the original dataset by T16a (\textit{left}) and the extracted subset (\textit{right}) are denoted in light blue. Results from LAAT are depicted in dark blue; \textit{Middle panel}: 3D representation of the results from EM3A on a zoomed-in perspective of the original dataset (\textit{left}) and its extracted subset \textit{(right}); \textit{Lower panel}: description is the same as that of the top panels, but now includes results from DimIndex, which is indicated in magenta.   }
               
         \label{LAAT_MBMS}
   \end{figure*}

\begin{figure*}
  \centering
   \includegraphics[width=\hsize]{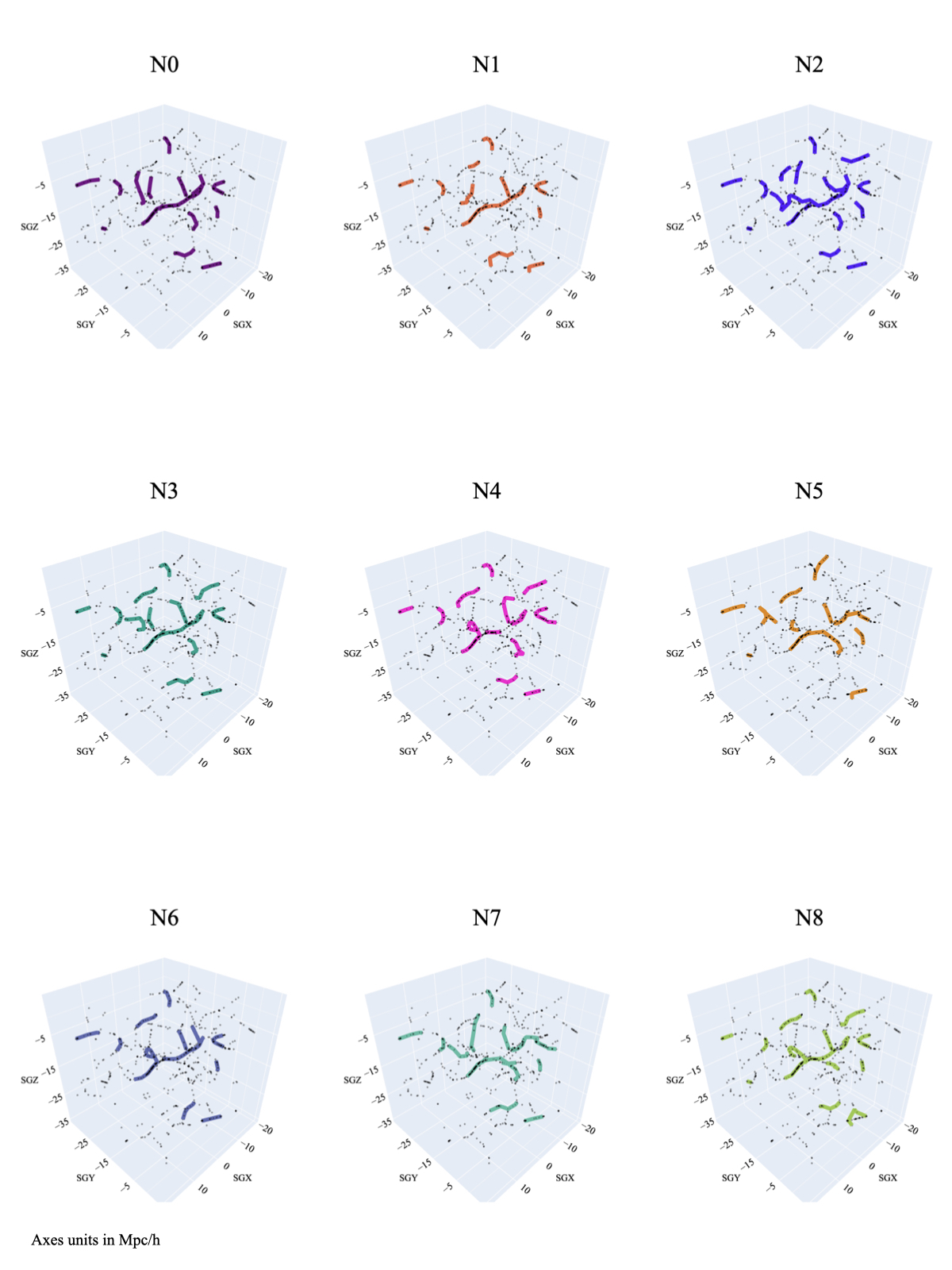}
     
      \caption{3D representation of the results from nine iterations of MMCrawling, continued from Fig.~\ref{Crawls_all}. }
               
         \label{Crawls_ite}
   \end{figure*}

In this section, we present a 3D visualisation of the results from 1-DREAM and tabulate the adopted input parameters in Table~\ref{tab:my_label}. Fig.~\ref{LAAT_MBMS} depicts the results from LAAT, EM3A, and DimIndex, followed by the results from MMCrawling in Fig.~\ref{Crawls_ite}. Table~\ref{tab:gal_pop} is a tabular presentation for Fig.~\ref{Morph}.

\begin{table}
    \caption{Input parameters for algorithms in 1-DREAM.}
    \centering
    \begin{tabular}{lcccl}
    \hline
    Algorithm. & Parameter & Value & Description  \\
    \hline
    \\
LAAT & $ r \in \mathbb{R}$ &3 &Neighbourhood radius  \\
& $\zeta \in \mathbb{R}$ & 0.05 &Evaporation rate \\
& $\kappa \in \mathbb{R}$ & 0.8 &Shape v. Pheromone\\
& $\omega  \in\mathbb{R}$ & 10 &Inverse temperature\\
&$ \gamma  \in \mathbb{R}$ & 0.05 &Deposited pheromone\\
& $F \in \mathbb{R}$ & 10 & Pheromone threshold\\
& $N_{epoch}  \in\mathbb{N}$ &100 &Number of Epochs\\
& $N_{steps} \in \mathbb{N} $ &1200 &Steps per epoch\\
& $N_{ants }\in \mathbb{N} $ &5 &Number of Agents\\
\hline
\\ 
EM3A+ & $r \in \mathbb{R}$ & 3  & Neighbourhood radius \\
     & $ \eta \in \mathbb{R}$ & 0.1 &  Learning rate \\ 
     & $ N_{s} \in \mathbb{N} $ &  100 &  Number of steps\\
    & $ N_{g} \in \mathbb{N}$ & 10 &  Number of generations \\
\hline
\\ 
DimIndex& $r \in \mathbb{R}$ & 3  & Neighbourhood radius&\\
     & $ \tau \in \mathbb{N} $ & 5 & Filtering threshold&\\ 
\hline
\\ 
MMCrawling& $r \in \mathbb{R}$ & 3  & Neighbourhood radius&\\
     & $ \beta \in \mathbb{R}$  & 0.5& Jumping tolerance\\

\hline
    \end{tabular}

    \label{tab:my_label}
\end{table}

\begin{table*}
\caption{Galaxy population in filaments around the Fornax-Eridanus Complex}              
\label{tab:gal_pop}      
\centering                                      
\begin{tabular}{ccccccc}        
\hline\hline    
Distance  & & $N_{gal}$ (ETGs) & & & $N_{gal}$ (LTGs) &   \\  
Mpc/$h$ &       &         &           &       &             & \\
 \hline   
& Total &  Groups &  Pristine &  Total &  Groups &  Pristine \\
 
(1) & & (2) & & & (3) &  \\
 \hline   
   \\
0.5 &      56 &       50 &          6 &     40 &      31 &         9 \\
1.0 &      39 &       34 &          5 &     94 &      64 &        30 \\
1.5 &      34 &       27 &          7 &     71 &      51 &        20 \\
2.0 &      17 &        9 &          8 &     62 &      34 &        28 \\
2.5 &       9 &        5 &          4 &     29 &      17 &        12 \\
3.0 &       7 &        5 &          2 &     21 &       7 &        14 \\
\hline
\hline
\\ 
\end{tabular}
\tablefoot{Column 1 -- Histogram bin; Column  2 -- Total number of ETGs in filaments, of which belong to groups/clusters, and pristine environments of filaments ; Column 3 --  Total number of LTGs in filaments, of which belong to groups/clusters, and pristine environments of filaments. }
\end{table*}

\end{appendix}

\end{document}